\documentclass[twocolumn,prx, superscriptaddress]{revtex4}
\pdfoutput=1

\usepackage[latin9]{inputenc}
\usepackage{mathrsfs}
\usepackage{amsmath}
\usepackage{amssymb}
\usepackage{graphicx}
\usepackage{comment}
\usepackage{placeins}
\usepackage{float}

\makeatletter

\providecommand{\tabularnewline}{\\}

\usepackage{mathrsfs}\usepackage{braket}
\usepackage{dsfont}
\usepackage{gensymb}
\usepackage{comment}
\usepackage{siunitx}
\usepackage{color}

\makeatother

\begin{document}
\begin{abstract}
We present a new method for the spectral characterization of pulsed twin beam sources in the high gain regime, using cascaded stimulated emission. We show an implementation of this method for a ppKTP spontaneous parametric down-conversion source generating up to 60 photon pairs per pulse, and demonstrate excellent agreement between our experiments and our theory. This work enables the complete and accurate experimental characterization of high gain effects in parametric down conversion, including self and cross-phase modulation. Moreover, our theory allows the exploration of designs with the goal of improving the specifications of twin beam sources for application in quantum information, computation, sampling, and metrology.
\end{abstract}
\title{Understanding high gain twin beam sources \\
 using cascaded stimulated emission}
\author{Gil Triginer\footnote{Equal contributors}}
\affiliation{Clarendon Labs, Department of Physics, Oxford University, Parks Road
	OX1 3PU Oxford}
\author{Mihai D. Vidrighin$^{*}$}
\affiliation{Clarendon Labs, Department of Physics, Oxford University, Parks Road
	OX1 3PU Oxford}
\author{Nicol\'as Quesada\footnote{Current affiliation: Xanadu, Toronto, Canada } }
\affiliation{Perimeter Institute for Theoretical Physics, Waterloo, ON, N2L 2Y5,
	Canada}
\author{Andreas Eckstein}
\affiliation{Clarendon Labs, Department of Physics, Oxford University, Parks Road
	OX1 3PU Oxford}
\author{Merritt Moore}
\affiliation{Clarendon Labs, Department of Physics, Oxford University, Parks Road
	OX1 3PU Oxford}
\author{W. Steven Kolthammer}
\affiliation{QOLS, Blackett Laboratory, Imperial College London, London SW7 2AZ, United Kingdom}
\author{J.E. Sipe}
\affiliation{Department of Physics, University of Toronto, Toronto, ON, M5S 1A7,
	Canada}
\author{Ian A. Walmsley}
\affiliation{Clarendon Labs, Department of Physics, Oxford University, Parks Road
	OX1 3PU Oxford}
\affiliation{Office	of	the	Provost	and	Department	of	Physics,	Imperial	College	London,	South	Kensington	Campus,	SW7	2AZ,	UK}

\maketitle
Many pivotal experiments in quantum optics and technology rely  on
twin beams generated by sources based on parametric nonlinear optical processes, and
in recent years there has been significant progress in the development
of such sources. In particular, high gain two-mode squeezing in modes with well-defined
spatial and spectral properties can be produced by parametric down-conversion
(PDC) in waveguiding structures, using quasi-phase-matching
and group velocity matching techniques \cite{Mosley08,Eckstein11,Helen16}. The generation of twin beam pulses with up to tens of photon pairs has been demonstrated experimentally \cite{Harder16}. Together with the development
of efficient photon-number-resolving detectors \cite{Fukuda11}, this
enables a range of new experiments in quantum optics, such as conditional
non-Gaussian state preparation \cite{Zavatta04,Wenger04} and boson sampling experiments \cite{lund2014boson,hamilton2017gaussian,lund2017exact,huh2015boson}.

In dispersion-engineered pulsed PDC sources, residual spectral
correlations between the down-converted beams, due to effects such as phase-matching
revivals and non-uniformity in the spectral phase of pump pulses \cite{Law00,Wasilewski06,Lvovsky07,Christ13},
result in the generation of a small number of independently squeezed spectral modes with the same spatial profile. In the high gain regime, self-phase modulation (SPM) of the pump pulses \cite{Sundheimer93} and cross-phase modulation (XPM) induced by the pump on the down-converted beams \cite{Liberale06} also affect the spectral structure of PDC emission.
In addition, time-ordering corrections to the commonly used perturbative description of PDC must be considered \cite{Christ11,Quesada14}.
The complex interplay between all these phenomena makes the study of pulsed PDC sources in the high gain regime challenging, and the development of new techniques for the experimental characterization and theoretical understanding of non-perturbative PDC a priority. This paper focuses on experimental characterization, while in a companion paper \cite{TheoryPaper} we  develop  theoretical approaches to treat the non-perturbative regime. 

A large body of work has already been devoted to the characterization of PDC sources, of course.  Stimulated emission tomography (SET) \cite{Liscidini13} is a general approach
used to infer the quantum properties of a spontaneous nonlinear process,
such as  spontaneous PDC, from an intensity measurement of its corresponding
stimulated process, in this example  difference frequency generation
(DFG). While it  has been the basis of many experiments \cite{Eckstein14,Fang14,Fang16,Rozema15,Jizan16,Ansari17},
and provides a detailed description of the spectral mode structure
of a PDC process, it  is generally not used to estimate the degree
of squeezing. Although  in principle it is possible to derive the
squeezing strength from the amount of parametric amplification on
a seed beam, in practice this is hindered by uncertainty in the amount
of overlap -- spatial, spectral, or polarization -- between the seeding and
amplified modes, as well as by optical loss.

An alternative approach consists in using higher order photon correlation
functions of the down-converted field to estimate the number of modes
in the source \cite{Christ11,Goldschmidt13,Burenkov17}. It  has the
advantage of being insensitive to loss, but provides little information
about the physical mechanisms that degrade the performance of the
source.

\begin{figure}[h]
\includegraphics[width=0.4\textwidth]{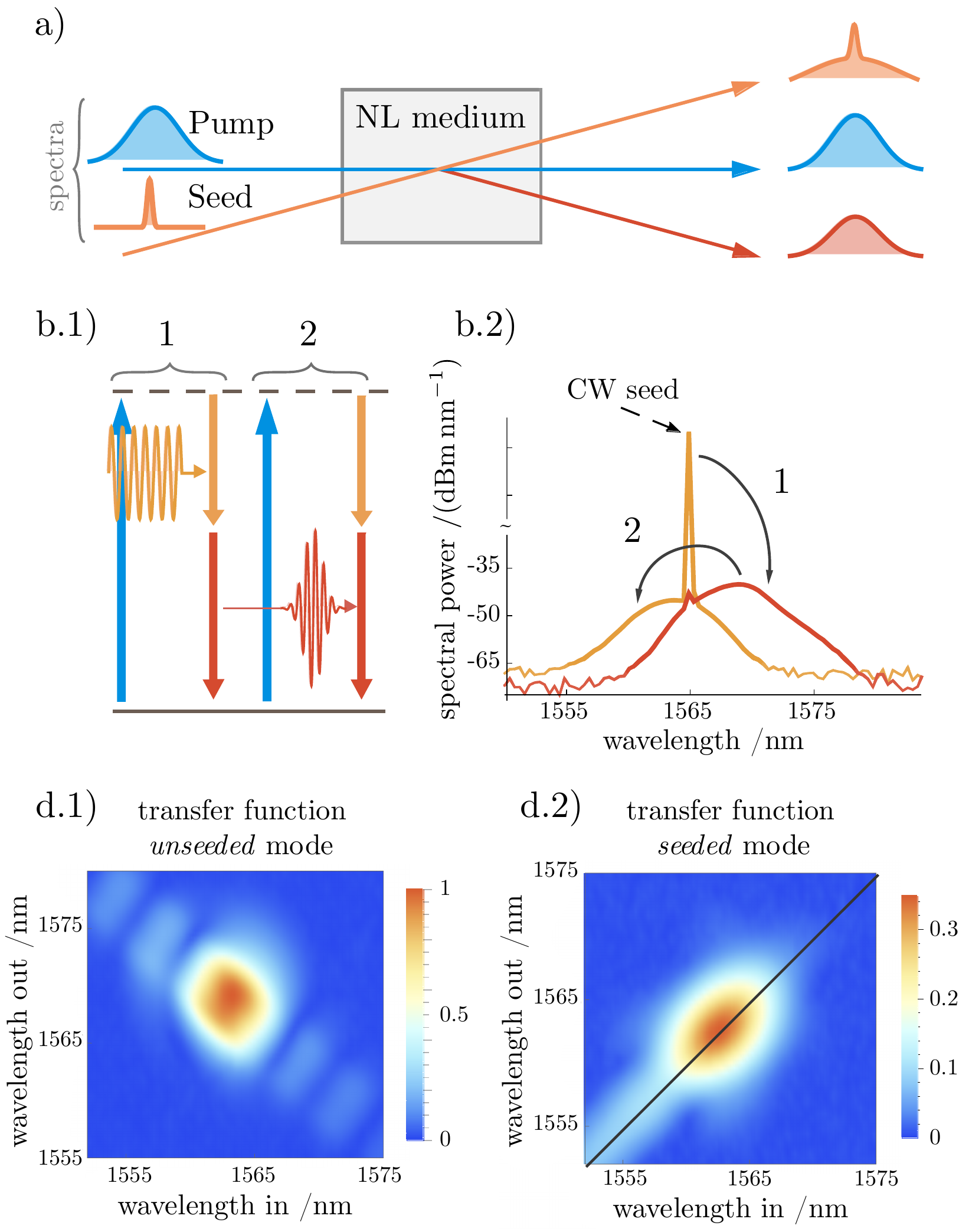}
\caption{a) Schematic of an extended SET experiment, including the broadband
light generated in the seeding beam through a cascaded DFG process.
b.1) Cascaded stimulated emission in a PDC crystal. Seeding one of the
two down-conversion modes with a continuous wave (CW) field (represented
in orange), bright light is generated through DFG between the pump
(represented in blue) and the seed. The DFG signal (represented in
red) acts as seed for a second DFG interaction with the pump field,
leading to broadband light generation in the mode that was originally
seeded. The process is repeated with decreasing strength. b.2) Measured
spectra of cascaded stimulated emission for a nearly degenerate type
II PDC source pumped with a pulse train centered at 780 nm. Seeding
with a CW laser at 1565 nm, we measured broadband signals in the opposite
polarization (red) as well as in the same polarization as the seed
(orange). c) The full spectral response of the PDC process reconstructed
by scanning the seed wavelength. The CW seed contribution was subtracted
from the spectra. The spectral responses are normalized to the maximum of the TF corresponding to the unseeded mode.}
\label{principle_of_operation}
\end{figure}

Here we demonstrate a method that provides detailed  spectral information
about the generated nonclassical light, and is applicable in  the
high gain regime. It extends SET to the measurement of all  emission processes resulting from a second-order optical nonlinearity. Two parametric amplification processes are seeded simultaneously, with the second seeded by the output of the first (see Figure \ref{principle_of_operation}), and so the resulting output intensities scale differently with squeezing strength.  The ratio between the output intensities can then be used to estimate squeezing parameters, without any knowledge of detection efficiencies or  mode overlaps, in a strategy that resembles self-referenced methods for estimating absolute detector efficiency using light generation \cite{Klyshko80,Lunghi14}. Additionally, the output of the second of the processes -- cascaded stimulated emission --  provides phase information due to the coherent addition of light generated by different pump frequencies, and  can be used to estimate the joint spectral phase of the  source.

To understand our cascaded stimulated emission measurements, we use simulations of the PDC process, with a small number of parameters
 extracted from experiments mainly in the low gain regime. We rely on Heisenberg-like equations of motion describing the spatial evolution of field amplitudes in the PDC source. Within the undepleted classical pump approximation,
we introduce an efficient method for solving these equations with arbitrarily
high gain. We find  excellent agreement between our experiments and our
theoretical predictions.
We can  identify  and characterize several effects in the large gain
regime that have been difficult to isolate before, including SPM of the pump field and XPM between
the pump and the down-converted modes. As our equations  describe
the PDC source at the quantum level, we are able to accurately quantify
the effects of these processes on the generated squeezed vacuum. We
also show how to use our experimentally validated simulations to optimize parameters of the
source with the goal of improving figures of merit, such as spectral purity. This
demonstrates that out approach is fruitful not only for understanding pulsed squeezed light sources, but also for optimizing their design. 

The outline of this paper is as follows: In Section I we introduce
the spectral transfer functions that can be used to characterize pulsed squeezing,
and discuss how they can be extracted from cascaded stimulated emission. Then we
give a theoretical framework for calculating parametric
down-conversion, valid both in the limits of small and
large gain, and discuss how most parameters for our theoretical description are extracted from
experiments in the small gain regime. In Section II we compare theory and experiment in both
the small and large gain limits. We discuss our results, and look
forward to future experiments, in Section III. Further technical details are
presented in the Appendices. 

\section{spectral description of pulsed twin beams}

\subsection{Input-output relations}

Assuming negligible propagation losses, and within the undepleted, classical
pump approximation, broadband two-mode squeezing in a waveguide
source supporting a single spatial mode for each field is fully
described by frequency-resolved input-output relations \cite{Christ13,Lvovsky07,TheoryPaper}:

\begin{subequations}
	\label{transferFunctions}
	\begin{align}
	a_{\mathrm{s}}^{(\mathrm{out})}(\omega)= & \int\!d\omega'\,U^{\mathrm{s},\mathrm{s}}(\omega,\omega')\,a_{\mathrm{s}}^{(\mathrm{in})}(\omega')\\
	& +\int\!d\omega'\,U^{\mathrm{s},\mathrm{i}}(\omega,\omega')\,a_{\mathrm{i}}^{\dagger(\mathrm{in})}(\omega'),\nonumber \\
	a_{\mathrm{i}}^{(\mathrm{out})}(\omega)= & \int\!d\omega'\,U^{\mathrm{i},\mathrm{i}}(\omega,\omega')\,a_{\mathrm{i}}^{(\mathrm{in})}(\omega')\\
	& +\int\!d\omega'\,U^{\mathrm{i},\mathrm{s}}(\omega,\omega')\,a_{\mathrm{s}}^{\dagger(\mathrm{in})}(\omega').\nonumber 
	\end{align}
\end{subequations}

Here $a_{\mathrm{s}(\mathrm{i})}^{(\text{in})}(\omega)$ denotes
the annihilation operator for the input signal (idler) field at frequency
$\omega$, and $a_{\mathrm{s}(\mathrm{i})}^{(\text{out})}(\omega)$
denotes the annihilation operator for the respective output mode.
The integrals run over the relevant bandwidths. Both input and output
operators satisfy the canonical commutation relations $[a_{\rm x}(\omega),a_{\rm y}^{\dagger}(\omega')]=\delta _{\rm x,y} \delta(\omega-\omega')$
and $[a_{\rm x}(\omega),a_{\rm y}(\omega')]=0$ with ${\rm x,y} \in \{{\rm s,i}\}$. The assumptions made
in order to derive these relations are detailed in our companion paper \cite{TheoryPaper}.

The transfer functions (TFs) $U^{\mathrm{s},\mathrm{s}},U^{\mathrm{s},\mathrm{i}},U^{\mathrm{i},\mathrm{s}}$,
and $U^{\mathrm{i},\mathrm{i}}$ completely describe the spectral
properties of the state generated in the PDC process, and all observable
quantities, such as correlation functions, can be predicted using
Equation (\ref{transferFunctions}). The cross-mode TFs $U^{\mathrm{s},\mathrm{i}(\mathrm{i},\mathrm{s})}$
are sufficient for describing spontaneous PDC, giving the joint spectral
amplitude of a photon pair in the low squeezing regime. However, in this work, we show that measuring the same-mode TFs $U^{\mathrm{s},\mathrm{s}(\mathrm{i},\mathrm{i})}$, which are only accessible in the stimulated regime, provides a redundancy that is very useful  for studying  experiments involving  high pump powers, where new effects such as SPM of the pump come into
play. In the application presented here, a full set of measurements
involving all the TFs  is  crucial for providing an explanation of the
 experimental results at high pump powers.

\subsection{Extracting spectral transfer functions from cascaded stimulated emission}
\label{extracting_TFs_from_DFG}
In SET measurements, a monochromatic seed field is coupled to one of
the two polarization-orthogonal down-conversion spatial modes. In the presence of the pump, this seed generates light in the other mode by parametric amplification, in a process  known as DFG. The emission is mapped over a range of seed frequencies,
resulting in a joint spectral distribution \cite{Liscidini13}. This has previously been used to measure the joint spectral intensity  of photon pair emission in the low squeezing regime \cite{Eckstein14}.

We can now describe SET using the notation introduced above. Equation (\ref{transferFunctions}) remains valid when the annihilation
and creation operators are replaced by classical field amplitudes
$\alpha_{\mathrm{s}(\mathrm{i})}(\omega)$ and their conjugates, as
we detail in Appendix \ref{appendix:classical_quantum_connection}.
Therefore, if the idler mode is seeded with the coherent amplitude
$\alpha_{i}^{\text{in}}(\omega)$, light will be generated in the
signal mode with the amplitude
\begin{align}
\alpha_{\mathrm{s}}^{(\mathrm{out})}(\omega)=\int\!d\omega'\,U^{\mathrm{s},\mathrm{i}}(\omega,\omega')\,\alpha_{\mathrm{i}}^{*(\mathrm{in})}(\omega').\label{seeding}
\end{align}
For large $\alpha_{i}^{\text{(in)}}$, the coherent component at
the output dominates spontaneous emission, and the measured power spectral density (PSD) in the signal mode
is $|\alpha_{s}^{\text{(out)}}(\omega)|^{2}$, with units of photon number per Hertz.  In Appendix \ref{perturbative_expansion} we show that, for low PDC gains, the amplitude of $U^{s,i (i,s)}$ scales linearly with a parametric gain proportional to the pump amplitude, as well as to the crystal nonlinearity. 

We can retrieve the absolute value of this transfer function using a narrowband
seed at frequency $\omega_{0}$, such that the PSD at the output is proportional to $|U^{\mathrm{s},\mathrm{i}}(\omega,\omega_{0})|^{2}$. The proportionality constant is given by the seed intensity, the detection
efficiency, and the overlap between the seed mode and the seeded down-conversion mode. Stacking the output spectra measured for a range of seed frequencies, a two-dimensional distribution proportional to $|U^{s,i}(\omega,\omega')|^2$ can be obtained.

According to the classical equivalent of  Equation (\ref{transferFunctions}), in the seeded measurement,
light is also generated in the same mode as the seed, with amplitude
\begin{align}
\alpha_{\mathrm{i}}^{(\mathrm{out})}(\omega)=\int\!d\omega'\,U^{\mathrm{i},\mathrm{i}}(\omega,\omega')\,\alpha_{\mathrm{i}}^{(\mathrm{in})}(\omega').\label{seeding2}
\end{align}

For low gain (see \cite{Wasilewski06} and Appendix \ref{perturbative_expansion}), $U^{\mathrm{i},\mathrm{i}}(\omega,\omega')\approx\delta(\omega-\omega')$,
and Equation (\ref{seeding2}) describes linear propagation, without frequency generation in the seed
mode. With a more intense pump field, the squeezing
strength increases, and a broadband ``pedestal'' is generated at
frequencies around the seed frequency. In Appendix \ref{perturbative_expansion}
we show that, in a series expansion of the TFs up to second order in the PDC interaction strength, the broadband contribution to the same-mode TF arises with the term proportional to the square of the parametric gain. A simple physical picture to understand the broadband pedestal is that of cascaded stimulated emission. The DFG field generated in the signal mode leads to the generation of a new DFG field in the idler mode (see Figure \ref{principle_of_operation}). 

In a manner analogous to the cross-mode transfer function reconstruction,  a two-dimensional distribution proportional to $|U^{\rm i,i}|^2$ can be obtained by measuring the spectral intensity in the idler mode for a range of idler seed frequencies. The two remaining transfer functions are obtained by scanning a narrowband seed in the signal mode and measuring the spectral intensities in the idler ($|U^{\rm i,s}|^2$) and signal ($|U^{\rm s,s}|^2$) modes. We denote the broadband part of a same-mode TF $U_{(\mathrm{b})}^{\mathrm{i},\mathrm{i}(\mathrm{s},\mathrm{s})}(\omega,\omega')=U^{\mathrm{i},\mathrm{i}(\mathrm{s},\mathrm{s})}(\omega,\omega')-\delta(\omega-\omega')$. To obtain this, we subtract the seed spectrum, which we measure in the absence of the PDC pump, from the output.

These seeded measurements are impacted by the (in general uncalibrated) coupling efficiency of the seed to the down-conversion mode,  $\eta_\text{in, s(i)}$, as well as by the detection efficiency of the generated outputs, $\eta_\text{out, s(i)}$. The stimulated emission from x to y where x,y $\in \{$s,i$\}$ is given by

\begin{align}
%\begin{split}
%\text{stim. emission from }x\text{ to }\text{y}&\propto
\eta_{\text{in},x}\,\eta_{\text{out},y}\,\left|U_{(\rm b)}^{\mathrm{y},\mathrm{x}}\right|^2,
%x, y &\in\{\rm s, i\}.
%\end{split}
\label{etas}
\end{align}
which hinders the retrieval of the absolute magnitude of the TFs, and therefore the inference of the parametric gain corresponding to the generated twin-beams. We rely on the different scaling of the same and cross-mode TFs with the parametric gain in order to extract it. We define a ratio of TF maxima,
\begin{align}
\kappa=\frac{\max[|U_{\text{(b)}}^{\text{s,s}}(\omega,\omega')|]}{\max[|U^{\text{s,i}}(\omega,\omega')|]}\frac{\max[|U_{\text{(b)}}^{\text{i,i}}(\omega,\omega')|]}{\max[|U^{\text{i,s}}(\omega,\omega')|]},
\label{ratio}
\end{align}
which can be obtained by replacing the TFs by the respective measured stimulated PSDs, as the $\eta_\text{in}$ and $\eta_\text{out}$ coefficients defined in equation \ref{etas} cancel out. As we have indicated previously, for low PDC gains, the amplitude of the broadband part of the same-mode TFs grows quadratically with the nonlinear gain, while the amplitude of the cross-mode TFs grows linearly: therefore the ratio $\kappa$ grows linearly with the parametric gain (in the low gain). As we will detail in subsequent sections, this different scaling allows us to obtain the PDC gain from $\kappa$, independently of the seeding and detection efficiencies, $\eta_\text{in, s(i)}$ and $\eta_\text{out, s(i)}$.

\begin{comment}
In order to fit the spectral distributions, we normalize both the measurements and the simulations to a maximum value of 1.
The phase of the transfer functions is not determined directly by seeded measurements. For instance, the spectral phase of the pump pulse cannot be extracted directly from standard (cross-mode) SET. However, we show in Section II that this phase has an effect on the absolute value of the same-mode TFs, and can therefore be determined from a complete cascaded stimulated emission measurement. XPM induced by the pump on the signal and idler is also easy to quantify by analyzing the same mode TFs.
\end{comment}

\subsection{Theory of parametric down-conversion}\label{theory}

We limit our theoretical treatment to waveguide sources with  a single transverse
spatial mode for each of the signal, idler, and pump fields. The spectral
structure of  PDC in such a scenario has been studied before in the
perturbative \cite{Grice97} and non-perturbative \cite{Lvovsky07,Wasilewski06,Christ13,Quesada14, quesada2015time}
regimes. We follow an approach close to that of  Wasilewski and
Lvovsky \cite{Lvovsky07,Wasilewski06}, describing the evolution of
fields in space rather than in time. At high pump powers, one expects
that SPM on the pump and XPM induced by the pump on the signal and idler must be considered, and indeed we find  these effects
significant in our KTP source. In our companion paper \cite{TheoryPaper}
we provide a  derivation of the equations of motion (EOMs) of the
signal and idler annihilation operators, with these effects included. In a medium where group velocity dispersion can be ignored within the bandwidth of the signal and idler modes, it is convenient
to use a frame of reference that propagates at the group velocity
of the pump beam \cite{TheoryPaper,MihaiThesis}. Assuming lossless propagation and an undepleted classical pump, the monochromatic annihilation operators for the slowly-varying envelopes of the signal(idler) in this propagating frame of reference, $a_{\mathrm{s(i)}}(z,\omega)$,
fulfill the following EOMs \cite{TheoryPaper,MihaiThesis}:

\begin{subequations}
 \label{eoms}
\begin{align}
\partial_{z}\, & a_{\mathrm{s}}(z,\omega)=i\,\Delta k_{\mathrm{s}}(\omega)\ a_{\mathrm{s}}(z,\omega)\nonumber\\
 & +i\,\gamma_{{\rm {PDC}}}\,g(z)\,(2\pi)^{-1/2}\!\int\!d\omega'\,\beta_{\mathrm{p}}(z,\omega+\omega')\,a_{\mathrm{i}}^{\dagger}(z,\omega')\nonumber\\
 & +i\,\gamma_{{\rm{XPM,s}}} (2\pi)^{-1} \int\!d\omega'\,\mathcal{E}_{\mathrm{p}}(\omega-\omega')\,a_{\mathrm{s}}(z,\omega'), \label{diffEqCompleteSignal} \\
\partial_{z}\, & a_{\mathrm{i}}(z,\omega)=i\,\Delta k_{\mathrm{i}}(\omega)\,a_{\mathrm{i}}(z,\omega)\nonumber\\
 & +i\,\gamma_{{\rm {PDC}}}\,g(z)\,(2\pi)^{-1/2}\!\int\!d\omega'\,\beta_{\mathrm{p}}(z,\omega+\omega')\,a_{\mathrm{s}}^{\dagger}(z,\omega')\nonumber\\
 & +i\,\gamma_{{\rm {XPM,i}}} (2\pi)^{-1} \int\!d\omega' \,\mathcal{E}_{\mathrm{p}}(\omega-\omega')\,a_{\mathrm{i}}(z,\omega'). \label{diffEqCompleteIdler}
\end{align}
\end{subequations}
The first term on the right hand side of the EOMs contains
$\Delta k_{\mathrm{s(i)}}(\omega)=(1/v_{\mathrm{s(i)}}-1/v_\mathrm{p}) (\omega-\bar{\omega}_{\mathrm{s(i)}})$. Here, $v_\mathrm{s(i)}$ and $v_\mathrm{p}$ are the group velocities of the signal(idler) and pump pulses, respectively and $\bar{\omega}_{\rm s(i)}$ is the central frequency of the signal (idler) mode. This term in the EOMs effectively describes linear propagation through the waveguide. 

The complex function $\beta_{\mathrm{p}}(z,\omega)$
is the pump spectral amplitude in the chosen reference frame, which evolves slowly along $z$ due to SPM inside the nonlinear region \cite{Shimizu67,TheoryPaper} , according to 
\begin{align}
	\partial_{z}\, \beta_{\mathrm{p}}(z,\omega)=\,i\,\gamma_{{\rm{SPM}}}\int\!d\omega'\,\mathcal{E}_{\mathrm{p}}(\omega-\omega')\,\beta_{\mathrm{p}}(z,\omega').
	\label{diffEqCompletePump}
\end{align}
The pump spectral amplitude satisfies the normalization
\begin{align}
\int\!d\omega\,|\beta_{\mathrm{p}}(z,\omega)|^{2}=\mathcal{E}_{{\rm {p}}}(0) = E_{{\rm {p}}},
\end{align}
where $E_{{\rm {p}}}$ is the energy in the pump pulse.
Note the difference in units between $\beta_{p}(z,\omega)$ and the $\alpha_{s(i)}(\omega)$ introduced earlier, near Equation (\ref{seeding},\ref{seeding2}). 

The constant $\gamma_{{\rm {PDC}}}$ in equations (\ref{diffEqCompleteSignal},\ref{diffEqCompleteIdler})
is the second order nonlinear coupling strength, and the function $g(z)$ accounts for the possibility of a sign reversal of this coefficient as is the case in periodically poled crystals. It  takes the values $1$ or $-1$ over the length of a periodically poled region, and
zero outside the crystal. 

The third term describes XPM
between the pump and the signal (idler), with $\gamma_{{\rm {XPM,s(i)}}}$
a coupling strength that can be different for the signal and idler.
This term also contains the frequency autocorrelation function of the pump spectral amplitude,
which in this frame of reference is spatially invariant \cite{TheoryPaper},
\begin{align}
\mathcal{E}_{{\rm {p}}}(\Delta\omega)=\!\int d\omega''\ \beta_{\text{p}}(z_0, \, \omega''-\Delta\omega)^{*}\,\beta_{\text{p}}(z_0, \, \omega'').
\end{align}
where $z_0$ is the input face of the waveguide.

The solutions of the EOMs \eqref{diffEqCompleteSignal} and \eqref{diffEqCompleteIdler} have a special form, which is related to the fact that the commutation relations are preserved \cite{TheoryPaper}. The TFs can always be written as:
\begin{subequations}
	\begin{align}
	U^{\text{s},\text{s}}(\omega,\omega') &= \sum_l \cosh(r_l) \, \rho_{\text{s}}^{(l)}(\omega)\tau_{\text{s}}^{(l)}(\omega'), \\
	U^{\text{s},\text{i}}(\omega,\omega') &= \sum_l \sinh(r_l)\, \rho_{\text{s}}^{(l)}(\omega)[\tau_i^{(l)}(\omega')]^*,\\
	U^{\text{i},\text{i}}(\omega,\omega') &= \sum_l \cosh(r_l)\, \rho_{\text{i}}^{(l)}(\omega)\tau_{\text{i}}^{(l)}(\omega'),\\
	U^{\text{i},\text{s}}(\omega,\omega') &= \sum_l \sinh(r_l)\, \rho_{\text{i}}^{(l)}(\omega)[\tau_{\text{s}}^{(l)}(\omega')]^*.
	\label{schmidt_decomposition}
	\end{align}
\end{subequations}
Here, $\rho_{\text{s}(i)}^{(l)}$ are normalized signal(idler) output spectral modes, $\tau_{\text{s}(\text{i})}^{(l)}$ are the corresponding input modes, and $r_l$ are the corresponding squeezing parameters.

Remarkably, if  SPM can be neglected and the PDC nonlinearity is uniform or has a homogeneous periodic poling, the EOMs (\ref{diffEqCompleteSignal}, \ref{diffEqCompleteIdler})
can be efficiently solved by exponentiating the discretized differential
operator \cite{TheoryPaper}. Scenarios  involving slow spatial variations
of either the pump amplitude (through SPM) or the effective
nonlinearity in periodically poled crystals
require an additional step. In those situations  we split the crystal
into a number of sections along the propagation direction, within
each of which the pump amplitude can  be assumed uniform. The input-output relations are then found by sequentially applying the transformations for all sections. This procedure remains very efficient --for the discretized transformations, it amounts to matrix multiplication--, allowing us to perform parameter sweeps effectively  even in the high gain regime.

In treatments of squeezing it is sometimes assumed that one can neglect both the cross-phase modulation of the signal and idler by the pump, and the SPM of the pump itself. We refer to treatments that make this assumption as ``$\chi^{(2)}$ models". Treatments that, in addition, assume a weak quadratic nonlinearity will be referred to as ``perturbative $\chi^{(2)}$ models".
The more general description that our EOMs (\ref{diffEqCompleteSignal}, \ref{diffEqCompleteIdler})
provide goes beyond this. The solutions of the EOMs recover features of PDC that have been introduced before as results of time-ordering corrections \cite{Quesada14}, including an enhanced squeezing rate as a function of the pump power, and  the distortion (frequency broadening) of the Schmidt modes \cite{Quesada14,Christ13,TheoryPaper}. Moreover, they account for the spectral transformation of the twin-beams caused by SPM and XPM. In consequence, we refer to our description as a ``$\chi^{(2)}/\chi^{(3)}$ model".

In subsequent sections, we address how all the parameters appearing in the EOMs (\ref{diffEqCompleteSignal}, \ref{diffEqCompleteIdler},\ref{diffEqCompletePump}) can be accurately determined, and we experimentally confirm the predictions of the equations.  

\begin{table*}[ht]
	\begin{tabular}{|l|c|c|c|r}
		\hline 
		Parameter  & Symbol  & Value  & Method  & \tabularnewline
		\hline 
		Pump spectral amplitude  & $\left|\beta_{\mathrm{p}}(z_0 \, \omega_{\mathrm{p}})\right|$  & ---  & Optical spectrum analyzer  & \tabularnewline
		Pump spectral phase  & $\arg\left[\beta_{p}(z_0, \, \omega_{\mathrm{p}})\right]$  & ---  & SPIDER and Spectral interferometry  & \tabularnewline
		Group velocity mismatch \ \ & $1/v_{\mathrm{s}}-1/v_{\mathrm{p}}$  & 1.70 ps/cm & Angle and bandwidth of phase-matching profile  & \tabularnewline
		& $1/v_{\mathrm{i}}-1/v_{\mathrm{p}}$  & -1.02 ps/cm & + direct interferometric measurement of group delay & \tabularnewline
		PDC coupling  & $\gamma_{{\rm {PDC}}}$ & $\SI{28}{\per\watt\tothe{1/2}\per\meter}$  & Ratio of the cross-mode and same-mode TF amplitudes  & \tabularnewline
		XPM coupling & $\gamma_{\text{XPM,s}}$  & $\SI{0.16}{\per\watt\per\meter}$ & Asymmetry of the same-mode TFs  & \tabularnewline
		& $\gamma_{\text{XPM,i}}$  & $\SI{0.059}{\per\watt\per\meter}$  &  &
		\tabularnewline
		SPM coupling & $\gamma_{\text{SPM}}$  & \SI{0.56}{\per\watt\per\meter}  & Fitting of TFs in the high gain regime & \tabularnewline
		\hline
	\end{tabular}
	\caption{Model parameters.}
	\label{parameter_table}
\end{table*}

\subsection{Extracting physical parameters from experiments}

Our experimental demonstration employs a wave-guided ppKTP type II PDC source. The pump pulses with central wavelength of $\SI{783}{\nano\meter}$ and pulse duration of around 1 ps, propagate along the ordinary axis of the crystal (H polarization). They generate a signal field with
the central wavelength at $\SI{1563}{\nano\meter}$ in the H polarization, and an idler
field at $\SI{1569}{\nano\meter}$ in the V polarization. More details are provided in Appendix \ref{appendix_experimental_details}.

\begin{figure}[h]
	\includegraphics[width=0.5\textwidth]{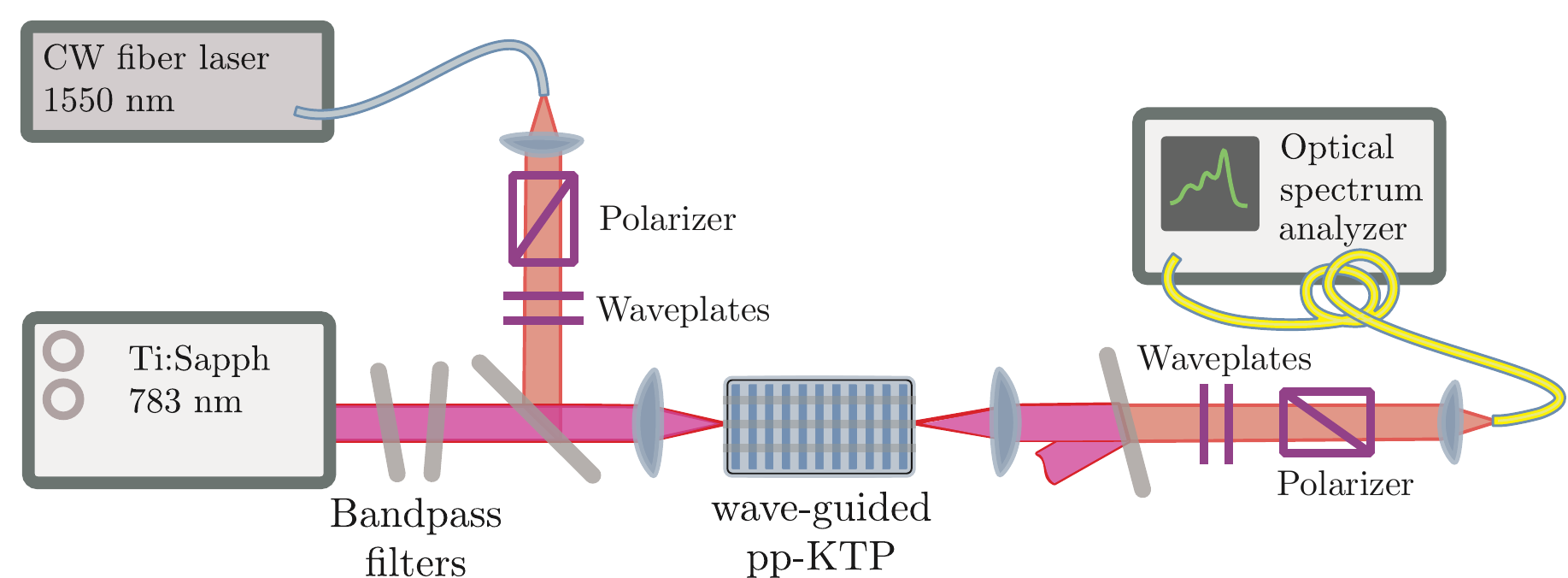}
	\caption{Experimental setup.}
	\label{setup}
\end{figure}

The design, reproduced from earlier work \cite{Eckstein11}, achieves a nearly separable joint spectrum. Therefore, the source generates almost single-mode broadband, orthogonally polarized twin beams. The waveguide confinement makes the source very bright. Previous to this work, up to 20 photons per pulse (on average) generated by spontaneous PDC has been reported \cite{Harder16}. However, the spectral structure of the twin beams in the high gain regime has not been systematically analysed. Here we show that we can correctly predict the absolute values of the measured TFs for this source up to an inferred mean photon number of 60 per pulse.

The cascaded stimulated emission setup is depicted in Figure \ref{setup}. We scan the frequency of a CW seed laser across the bandwidth of the signal/idler mode. After selecting a downconverted field by polarization filtering, we record the spectrum of the generated light using an optical spectrum analyzer (OSA). We performed measurements for pump pulse energies ranging
from 125 pJ to 600 pJ. For each pump power, we obtained all four TFs (absolute value) as explained in Section \ref{extracting_TFs_from_DFG}.

Most parameters in equations \ref{eoms} can be measured for this source in the low gain regime, either from stimulated emission data or from simple complementary experiments, as we summarize in this section. The values of the parameters, as well as the methods we used to obtain them, are listed in Table \ref{parameter_table}. The reader interested in replicating our method can find the necessary details in Appendix \ref{identifying_EOM_parameters}.

We measure the pump power spectral amplitude, $|\beta_\mathrm{p}(z = 0, \, \omega)|^2$ (Appendix \ref{parameters_pump_intensity}) with an optical spectrum analyzer (OSA). We extract the spectral phase of the pump, $\arg\big(\beta_\mathrm{p}(z=0, \, \omega)\big)$ using a standard ultrafast spectral phase measurement, SPIDER, for the laser pulse, and a spectral self-interference measurement (Appendix \ref{parameters_pump_phase}) for the phase added by the bandpass filters used to shape the pump pulse, as shown in Figure \ref{setup}. 

We obtain the group velocity mismatch between pump and signal(idler) pulses, $(1/v_{\mathrm{s(i)}}-1/v_\mathrm{p})$, from the angle and bandwidth of the phase-matching function in a low gain JSI (Appendix \ref{parameters_group_velocity_mismatch}). This is confirmed by a direct interferometric measurement of the differential delay between the two polarizations (signal/idler) using a broadband field that propagates through the birefringent ppKTP crystal (Appendix \ref{parameters_group_velocity_mismatch}).  

We extract the PDC interaction strength, $\gamma_\text{PDC}$, from the ratio between the maxima of the same-mode and cross-mode TFs in the low gain regime, where these amplitudes are not influenced by other nonlinear processes (Appendix \ref{parameters_PDC_gain}). The XPM interaction strength is obtained from the shape of the same-mode TF in the low gain regime alone, as we show in the next section, as well as in Appendix \ref{parameters_XPM}. The SPM interaction strength, $\gamma_\text{SPM}$, is the only parameter that we extract from a fit in the high gain regime: once all other parameters are fixed, $\gamma_\text{SPM}$ is chosen to optimize the fitting of the high gain TFs (Appendix \ref{parameters_SPM}).  

\section{Experimental demonstration}

\begin{figure*}
	\includegraphics[width=1\textwidth]{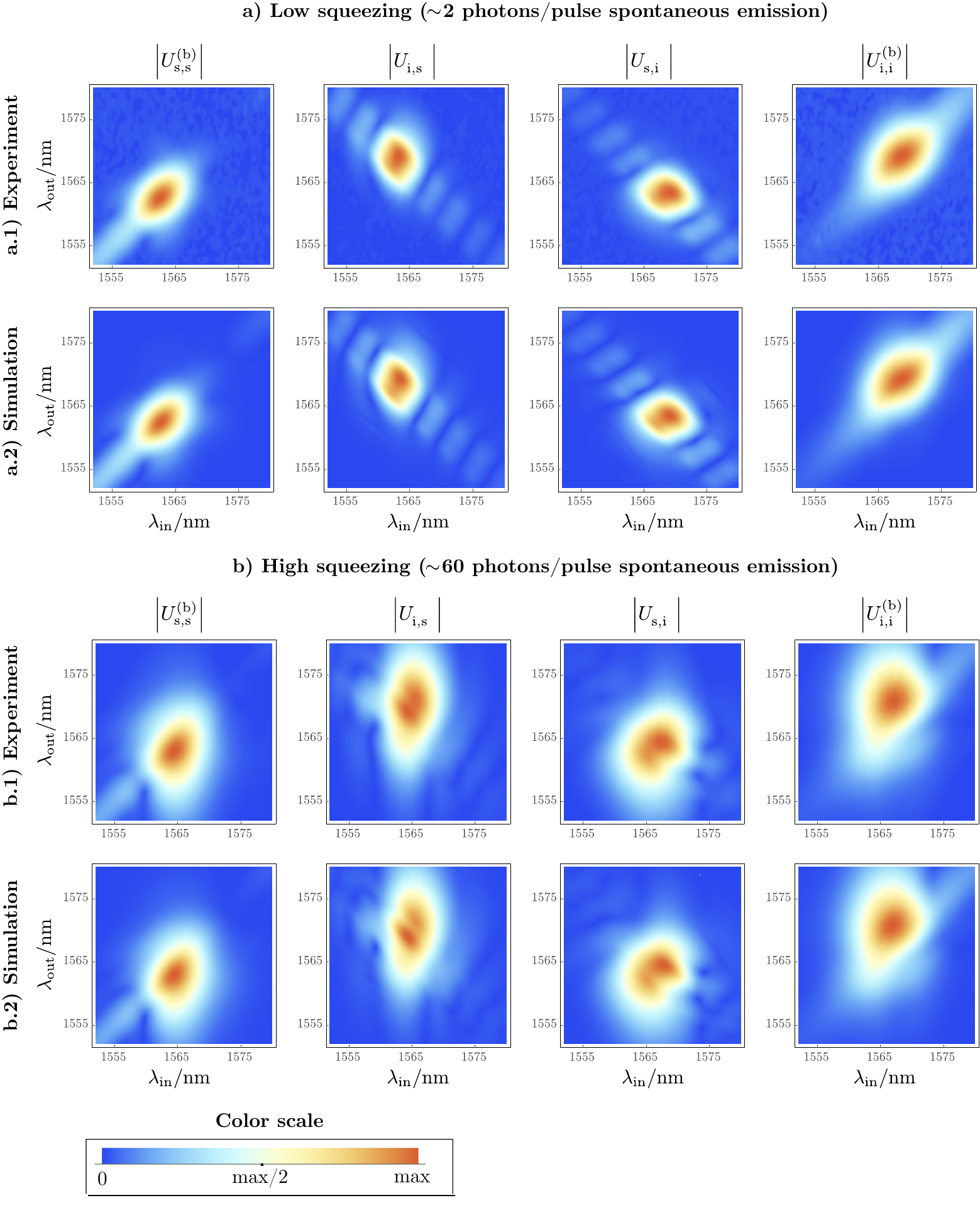}
	\caption{All measured and modeled TF amplitudes, in the low and high gain regimes, corresponding to strong and weak squeezing. The color scale is normalized separately for each density plot. The TF amplitudes are plotted as a function of wavelength instead of angular frequency,
		where these are connected by $\lambda=2\pi c/\omega$.}
	\label{big}
\end{figure*}

Here we report our cascaded stimulated emission measurements for an ample range of gains, and demonstrate the excellent agreement with our simulated TFs. Figure \ref{big} summarizes the results for the lowest and highest pump pulse energies. Each measured TF amplitude is compared to the prediction of our EOMs
(\ref{diffEqCompleteSignal},\ref{diffEqCompleteIdler}), with parameters fitted as explained in detail
in Appendix \ref{identifying_EOM_parameters}. Remarkably, a single model is able to accurately account for
all the experimental data.  In
what follows we report on the most important features of
the measured TFs, and provide their physical interpretation.

\label{sec:experimentalDemonstration}

\subsection{Low gain spectra}

We first measured the cross-mode stimulated emission using a relatively low pump power (125 pJ),
reproducing a standard SET experiment. The square roots of the resulting 2-dimensional PSDs, corresponding to the absolute value of $U^\text{s,i}$ and $U^\text{i,s}$, are depicted in the low gain section of Figure \ref{big}. We inferred a spontaneous emission rate of approximately 2 photons/pulse. In this regime, the cross-mode TF is equal to the joint spectral amplitude of the generated photon pairs.

In the same setup, we were able to resolve broadband stimulated
emission in the same polarization as the CW seed, centered
around the seed wavelength. By
scanning the seed wavelength, we reconstructed the two-dimensional spectral distributions, which are shown in the low gain section in Figure \ref{big}. As discussed in section \ref{extracting_TFs_from_DFG},
these are proportional to the absolute value of the same mode TFs $U^{\mathrm{s},\mathrm{s}}_{(\rm b)}$
and $U^{\mathrm{i},\mathrm{i}}_{(\rm b)}$.

\begin{figure}[h]
	\includegraphics[width=0.3\textwidth]{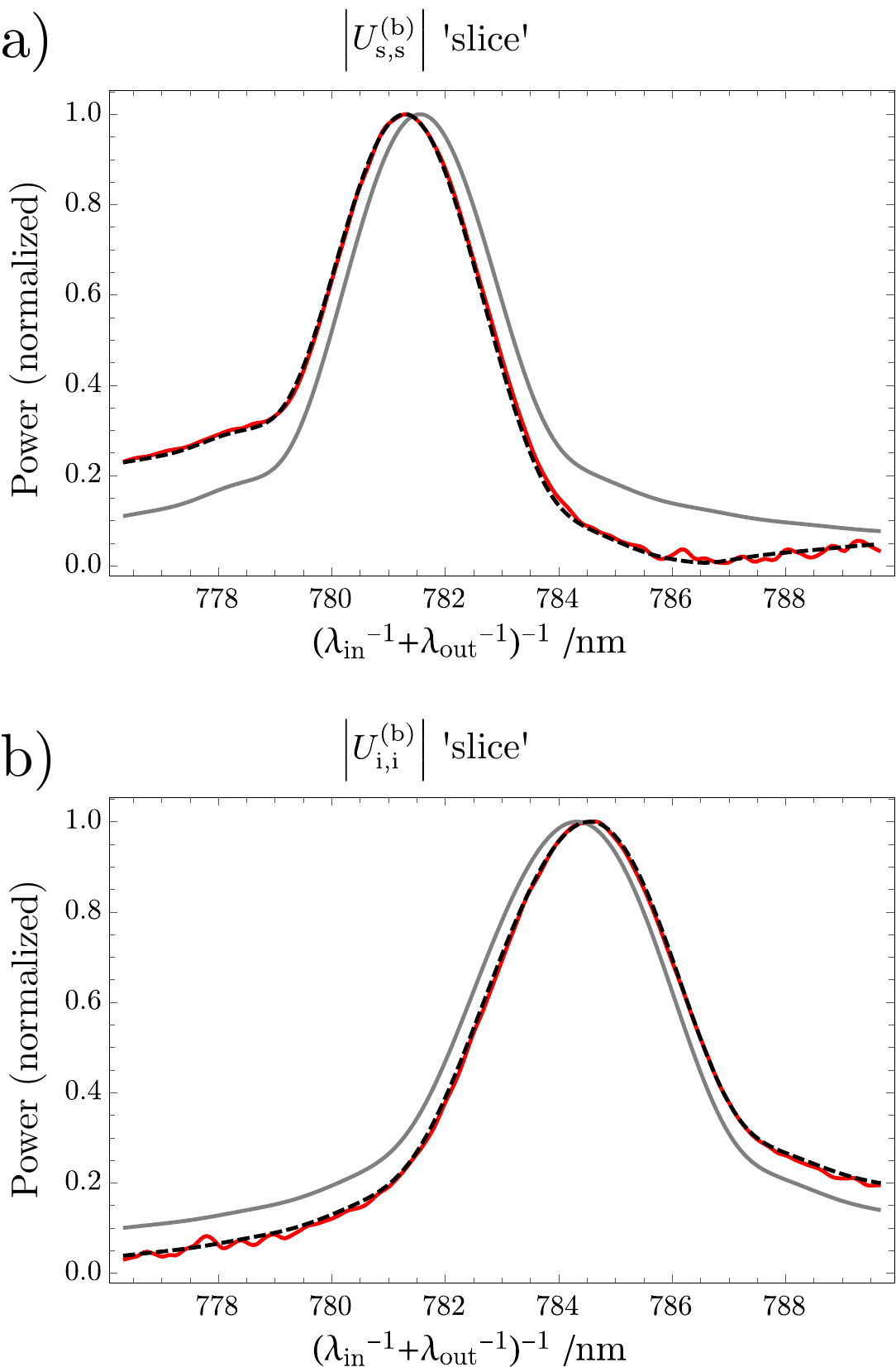}
	\caption{Same-mode TF slices showing the effect of XPM. The plots show $\left|U_{(\mathrm{b})}^{\mathrm{s},\mathrm{s}(\mathrm{i},\mathrm{i})}(\lambda_{\mathrm{in}},\lambda_{\mathrm{out}})\right|$
		for $\lambda_{\mathrm{in}}=\lambda_{\mathrm{out}}$. Red: Measurement
		(background subtracted). Black dashed: model prediction. Gray: prediction
		of a model that does not include XPM. The asymmetry arises from superposing
		the signals generated by cascaded stimulated PDC and XPM.}
	\label{XPMfitting} 
\end{figure}

In Figure 4, we compare experiment and simulation along the $\omega' \equiv  \omega_\text{in} = \omega_\text{out} \equiv \omega $ cut of the same mode TFs. We added for comparison the prediction of a $\chi^{(2)}$ model. The $\chi^{(2)}$ model is significantly less accurate, failing to predict the asymmetry in the tails of the distribution observed experimentally. We found that this asymmetry is explained by the presence of cross-phase modulation induced by the pump pulse on the down-converted modes. For frequencies far from the PDC phasematching, the effect of XPM appears as a pedestal around the CW seed. Along the $\omega_\text{in}=\omega_\text{out}$ contour, XPM results in a fixed background. The broadband component of the same-mode PDC TFs, as obtained from our simulations and analytically proven in Appendix \ref{perturbative_expansion}, shows a $\pi$ phase-jump around the phase-matching central wavelength. As a result, the coherent addition of the constant XPM background and the PDC signal results in an asymmetric shape, reminiscent of a Fano resonance. As indicated above and detailed in Appendix \ref{parameters_XPM}, we extracted the XPM interaction strength by optimizing the experiment-theory fit for the profiles in Figure 4. We found that the XPM interaction strength is not equal for the signal and idler modes, and we analyze this observation in detail in Appendix \ref{factorof3}. The signal (with the same polarization
as the pump) has an XPM interaction strength roughly three times greater
than the idler (with polarization orthogonal to that of the pump).
The asymmetry of the same-mode TFs is observed to be independent of the pump power, as both cascaded PDC and XPM scale
quadratically with the pump amplitude in the low gain regime. This makes it a robust feature, and a good way to estimate the ratio between $\chi^{(2)}$ and $\chi^{(3)}$ interaction strengths.

\subsection{High gain spectra}

Increasing the pump power, we observed additional deviations from
the $\chi^{(2)}$ model discussed above.  The high gain section in Figure \ref{big}  shows measurements with a pump pulse energy
of 600 pJ; we inferred a spontaneous generation rate of about 60 pairs of photons per pump pulse.  As expected in the high gain regime, phase matching lobes are suppressed, and the spectra are broadened \cite{Christ11}. In addition to these effects, accounted for by the $\chi^{(2)}$ model, the cross-mode TFs are highly distorted and the same-mode TFs show a
significant asymmetry with respect to the $\omega_\text{in}=\omega_\text{out}$ line.

\begin{figure}[h]
	\includegraphics[width=0.45\textwidth]{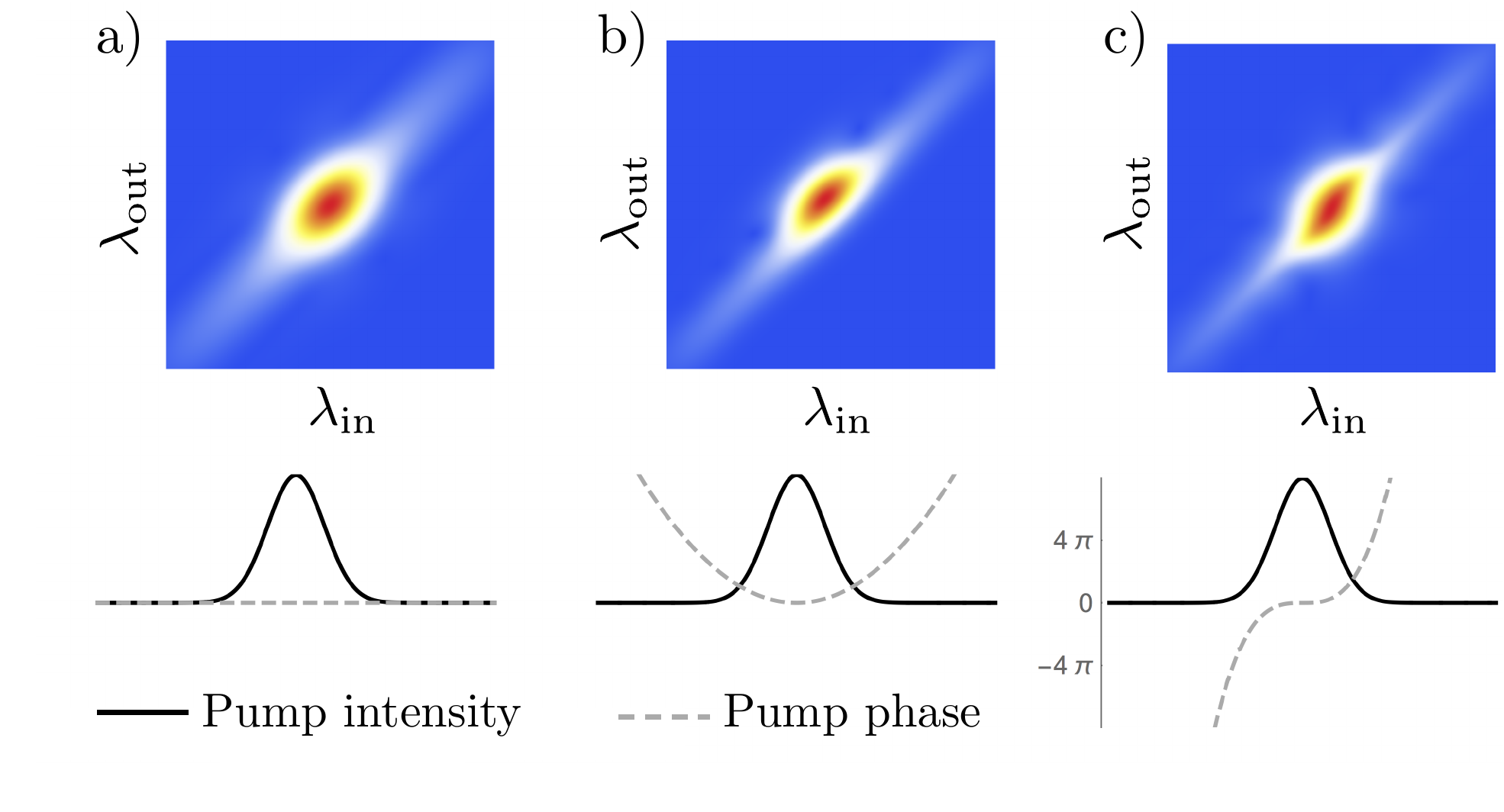}
	\caption{The amplitudes of the same-mode TFs, $U_{(\mathrm{b})}^{\mathrm{s},\mathrm{s}(\mathrm{i},\mathrm{i})}(\omega,\omega')$. Simulated same-mode TF for a) Fourier
		limited, b) chirped, and c) higher-order chirped Gaussian pump pulses
		(intensity and bandwidth in arbitrary units). A chirped pump causes a displacement of the same-mode TF from the diagonal (b), while a higher-order chirp leads to a more complicated deformation of the same-mode TF.}
	\label{spectralPhaseOnUss} 
\end{figure}

We can fully account for these features by  including the effect of self-phase modulation of the pump into Equations (\ref{diffEqCompleteSignal},\ref{diffEqCompleteIdler}). The distortion of the cross-mode TFs is related to the broadening and splitting of the pump spectrum through SPM. The shift of the same-mode TFs can be explained by the chirp accumulated by the pump pulse, also due to SPM. As illustrated in Figure \ref{spectralPhaseOnUss}, the spectral
phase of the pump is mapped onto the absolute value of the same-mode TFs, as signals generated by different pump frequencies add up coherently in the cascaded frequency
generation process. In particular, a quadratic phase in the pump field causes a shift of the same-mode TFs away from the $\omega_\text{in} = \omega_\text{out}$ line. The excellent agreement between this full model and the high gain data is clear evidence that SPM becomes relevant at this pump pulse energy.

\subsection{Scaling and absolute, loss-independent validation}

\label{scaling_and_validation} 
\begin{figure}[h]
	\includegraphics[width=0.3\textwidth]{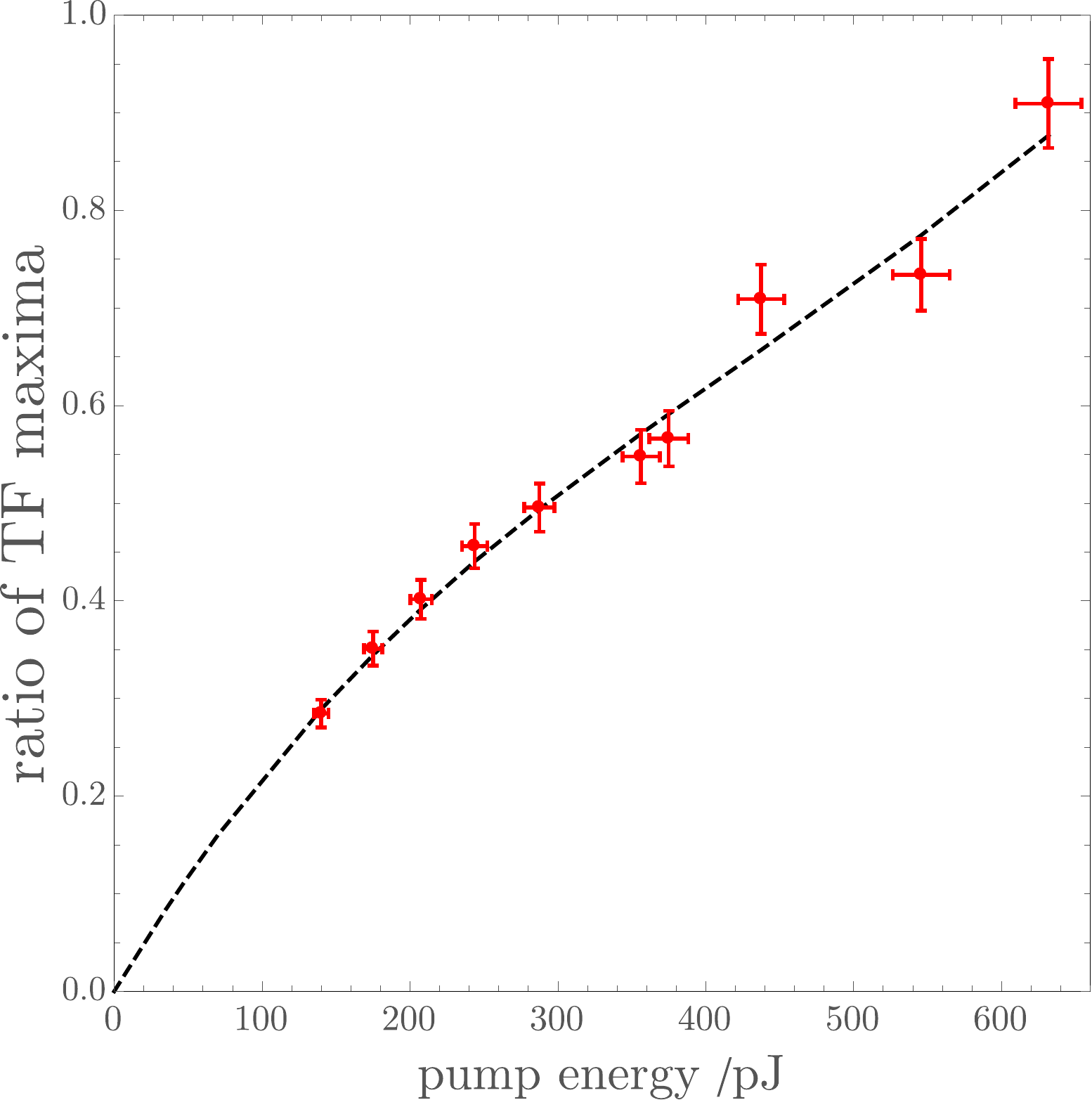}
	\caption{Ratio between the self and cross-mode transfer functions ($\kappa$)
		for different pump pulse energies.}
	\label{ratiosFigure} 
\end{figure}

The different scaling of same- and cross-mode TFs gives us access to the twin-beam gain using cascaded stimulated emission measurements outside of the high gain regime. As we show in Section \ref{extracting_TFs_from_DFG} and Appendix \ref{parameters_PDC_gain}, we can use ratios of the measured PSDs to construct the coefficient $\kappa$, defined in Equation \ref{ratio}, which grows monotonically with the PDC interaction strength, and which is independent of detection efficiency and mode overlap. Note that this coefficient captures features related to the gain of the process in both the low- and high gain regimes, and that in the low gain it gives information pertaining to the squeezing parameters characterizing the transformation of operators in equation \eqref{schmidt_decomposition}.  Figure \ref{ratiosFigure} shows that our theory correctly predicts the scaling of $\kappa$ observed in the experiments.

\subsection{Difference between measurement and simulation}

\begin{figure}[h]
	\includegraphics[width=0.35\textwidth]{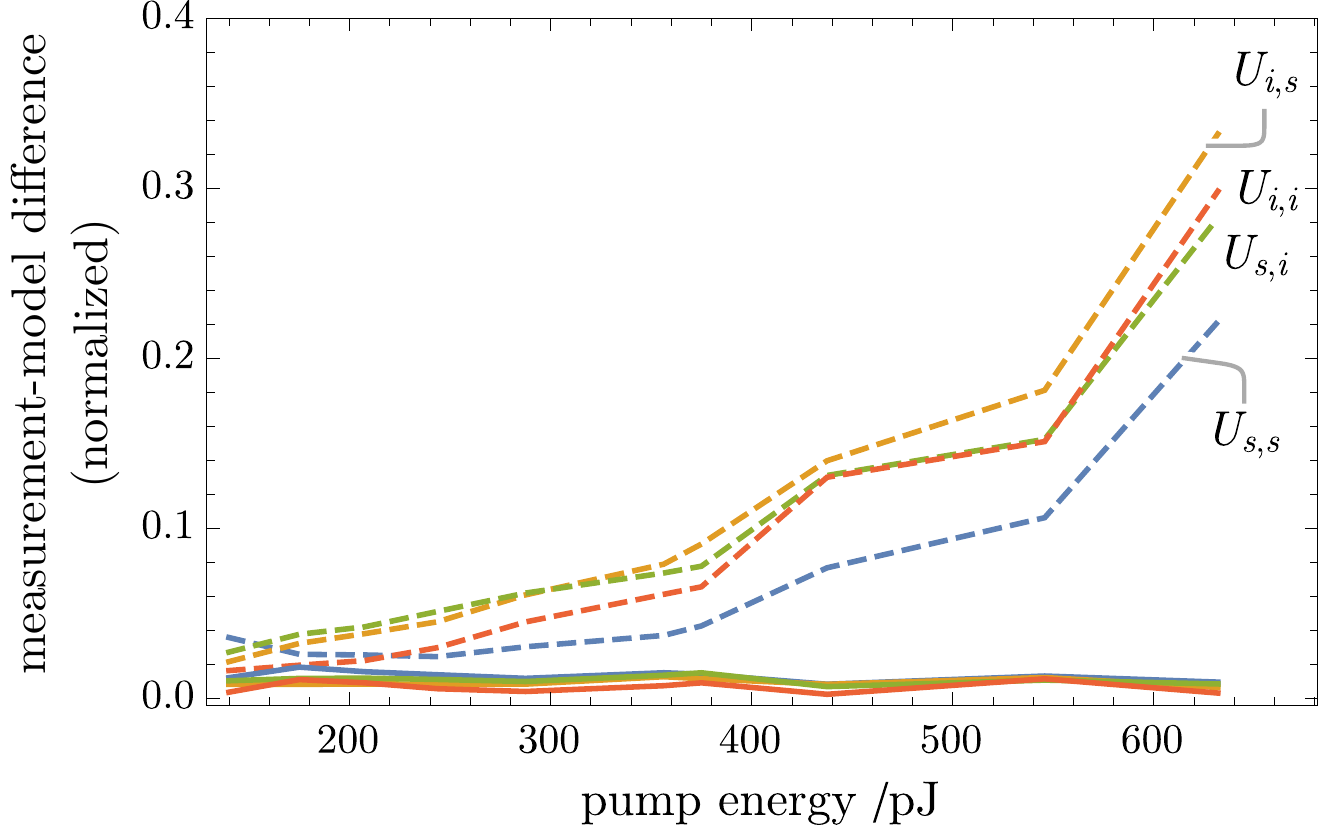}
	\caption{Mean-squared error between measured and modeled TFs (details in the
		Methods section). Solid lines -- $\chi^{(2)}/\chi^{(3)}$ model; dashed lines --
		$\chi^{(2)}$ model.}
	\label{errorsFigure} 
\end{figure}

Our theoretical model is consistent with experimental data for a broad range of spontaneous photon pair generation rates, from 1 photon pair/pulse on average to 60 photon pairs/pulse. To quantify the agreement between experiment and theory, we introduce the integrated squared error between the measured and simulated absolute values of the TFs, normalized to unit area:
\begin{align}
\varepsilon^{\rm x,y}&= \int\!d\omega \, d\omega' \left|  \frac{|U^{\rm x,y}_\text{meas (b)}(\omega,\omega')|}{\mathcal{N}_\text{meas}^{\rm x,y}} - \frac{|U^{\rm x,y}_\text{sim (b)}(\omega,\omega')|}{\mathcal{N}_\text{sim}^{\rm x,y}}  \right|^2\!\!,\nonumber \\
&\mathcal{N}_\text{sim}^{\rm x,y} = \left[\int\!d\omega \, d\omega' \, \left|U_\text{meas (b)}^{\rm x,y}(\omega,\omega')\right|^2\right]^{1/2},\nonumber \\
&\mathcal{N}_\text{sim}^{\rm x,y} = \left[\int\!d\omega \, d\omega' \,\left|U_\text{sim (b)}^{\rm x,y}(\omega,\omega')\right|^2\right]^{1/2},\nonumber \\
&{ \rm x, y} \in\{\rm s, i\}.
\label{error_metric}
\end{align}
Here, $U^{\rm x,y}_\text{meas (b)}$ is the broadband part of the power spectral density measured in mode $y$ when seeding mode $x$, and  $U^{\rm x,y}_\text{sim (b)}$ is the corresponding simulated TF. In Figure \ref{errorsFigure} we show that, using the calibrated $\chi^{(2)}/\chi^{(3)}$ model (solid lines), $\varepsilon^{\rm x,y}<2\%$ for all pump powers. On the other hand, neglecting the effect of XPM and SPM (dashed lines) we cannot produce a model that fits the experimental data for all pump powers. 
\begin{comment}
The third order nonlinearity estimate extracted here is in agreement with the literature, which reports $n_2 = 2.4 \times 10^{-15} \, \text{cm}^2/\text{W}$ for KTP crystals \cite{DeSalvo93, Sundheimer93}. Note that $\gamma_\text{SPM}=\frac{2\pi}{\lambda}n_2 A_\text{eff}^{-1}$ and our weakly guided waveguides have a cross-section of approximately $4 \mu\text{m} \times 4 \mu\text{m}$ \cite{Eckstein11}. 
\end{comment}

\section{Performance of the source}

We have characterized a source designed to provide  a high degree of squeezing in a small number of spectral modes. We have developed
a theoretical description that is validated by our experimental results. We now use this description to analyze the performance of the source. In particular, we discuss high gain effects on brightness and spectral purity.

\begin{figure}[h]
\includegraphics[width=0.35\textwidth]{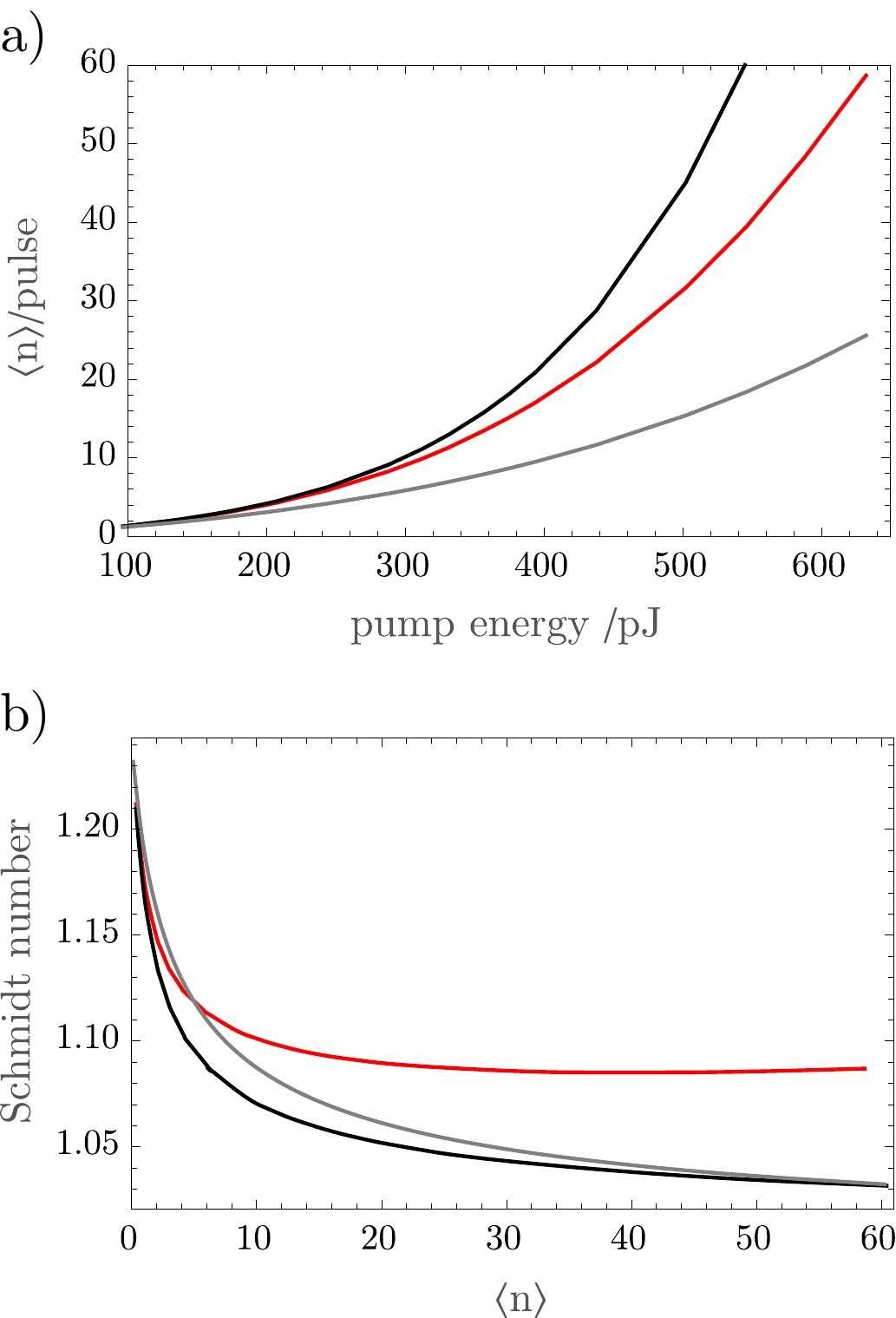}
\caption{a) Number of spontaneously generated photons as a function
of pump pulse energy. b) Spectral Schmidt number as a function of
the number of spontaneously generated photons. Red -- the $\chi^{(2)}/\chi^{(3)}$ model; black -- a $\chi^{(2)}$ model (only PDC, no SPM or XPM); gray -- perturbative $\chi^{(2)}$ model (squeezing parameters $r_l = \sqrt{P} \, c_l$) with the same predictions in the low gain regime.}
\label{schmidtNumberFigure} 
\end{figure}

\subsection{Photon number}
We can calculate the spontaneously generated mean photon number by adding the squares of the
Schmidt coefficients of the cross-mode TF \cite{Christ13}, $\langle n \rangle = \sum_l \sinh^2(r_l)$. A perturbative computation of the cross-mode TFs (e.g. Appendix \ref{perturbative_expansion}) yields squeezing parameters linearly proportional to the pump amplitude, and thus a spontaneously generated mean photon number $\langle n \rangle = \sum_l \sinh^2(c_l \sqrt{{E}_p})$, where ${E}_p$ is the pump pulse energy, and $c_l$ are constants. Our theoretical approach based on integrating EOMs offers revised predictions, previously treated as time-ordering corrections  \cite{Christ13, Quesada14}. For $\langle n\rangle>2$, the squeezing parameters scale nonlinearly with the pump amplitude, leading to an enhanced growth of $\langle n \rangle$. On the other hand, SPM of the pump pulse leads to a lower $\langle n\rangle$ than that predicted by a $\chi^{(2)}$ model. Figure \ref{schmidtNumberFigure}.a shows our results: we infer $\langle n \rangle$ from the fitted TFs and find that we are well into the regime where non-perturbative corrections become significant, and that SPM also has a significant effect on the spontaneously generated photon number. Our simulations show that the effect of XPM on the mean photon number is negligible.

\subsection{Spectral purity}
We now consider the spectral multimodeness of the source, quantified by the Schmidt number, defined as
$K=(\sum_{l}\sinh(r_{l})^{2})^{2}/(\sum_{l}\sinh(r_{l})^{4})$
\cite{Eberly06,Christ11}. A low Schmidt number implies that there is little correlation between the frequency spread of the signal and idler modes, and thus, ultimately, very low spectral entanglement. This situation therefore allows high purity single photons
to be heralded from pairs generated in the low gain regime. More generally, the Schmidt
number quantifies the maximum visibility of the second order interference between
the generated light and another optical mode \cite{Iskhakov13}. It has been shown that the Schmidt number of a PDC source is reduced in the high gain regime, due to the dominance of strongly populated modes \cite{Christ13,Dyakonov15}. We found that the pump SPM counteracts this effect: the Schmidt number saturates at a moderate pump power. Figure \ref{schmidtNumberFigure} illustrates these observations. Our simulations show that the effect of XPM on the spectral Schmidt number is negligible.

\begin{figure}[h]
	\includegraphics[scale=0.45]{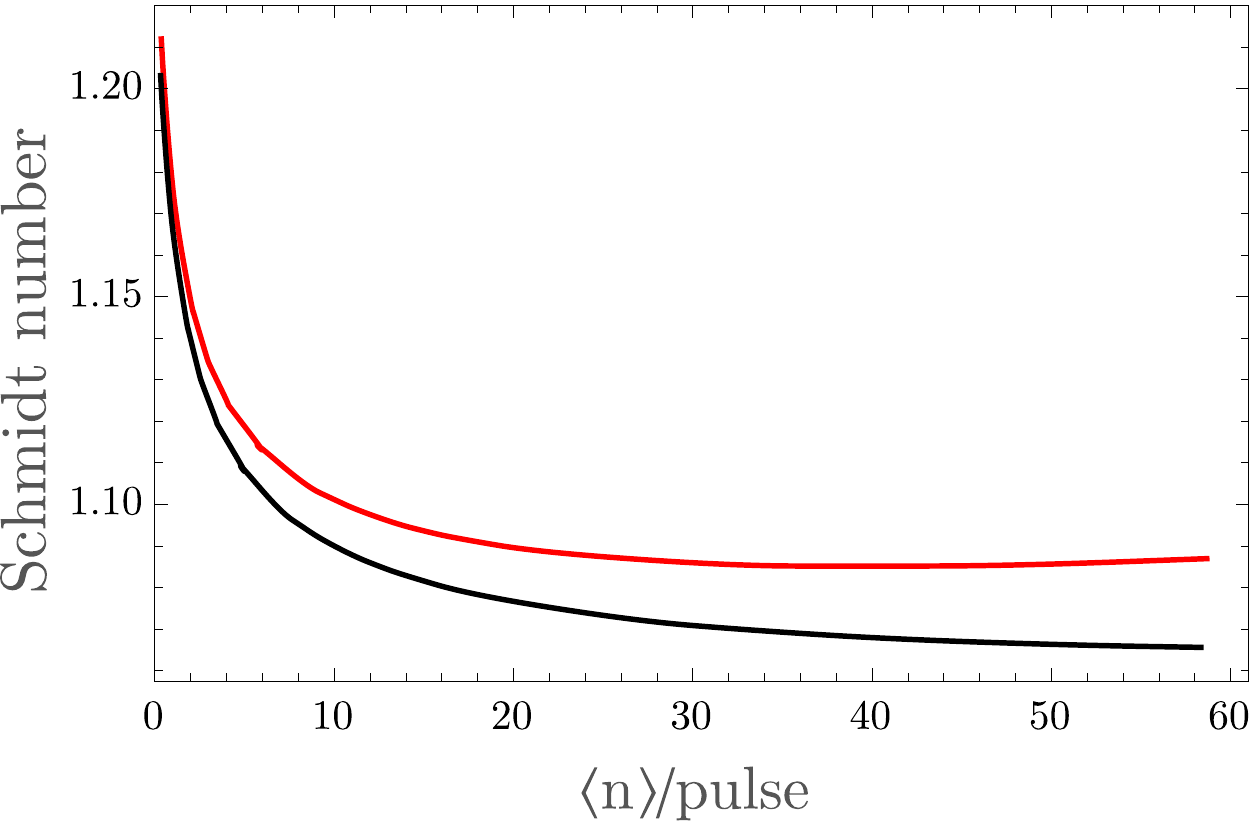}
	\caption{Effect of pre-chirping pump pulses on the spectral Schmidt number
		of a PDC source as a function of the average number of photons generated.
		Red -- Fourier-limited pump pulses; black -- pre-chirped pump pulses.}
	\label{pre_chirped_pump} 
\end{figure}

\subsection{Improving source performance}

We have shown that SPM due to the $\chi^{(3)}$
nonlinearity of KTP has an impact on the performance of our source in the high gain regime. The results of our study, however, suggest a simple optimization to counteract this effect: pre-chirping of the pump pulses. If the sign and magnitude
of the chirp are appropriately chosen, it is possible to counter the reduction in purity and brightness. In Figure \ref{pre_chirped_pump}
we show the Schmidt number
predicted by our theory as a function of the spontaneously generated photon number, contrasted with what could be achieved using chirped pump pulses. The dispersion parameter required to optimize the performance of the source is $D=178\ \text{fs/nm}$, which can be easily achieved using a grating-based pulse compressor.

\section{Conclusions}

In this work we have demonstrated a general experimental framework to characterize broadband twin-beam sources in the high gain regime. We have introduced cascaded stimulated emission tomography, a seeded measurement which generalizes SET, providing additional spectral information about the generated twin-beams. In particular, we have shown that the redundancy offered by this measurement allows for the self-referenced inference of the PDC interaction strength, independently of seeding and detection efficiency. Using the information offered by cascaded SET, together with a small number of complementary measurements, we have fitted a theoretical model of twin-beam generation, which yields all the complex TFs that describe the process. We have experimentally identified and quantified non-perturbative, SPM, and XPM effects in a high gain PDC source. In fact, the development of the EOMs (\ref{diffEqCompleteSignal}, \ref{diffEqCompleteIdler}, \ref{diffEqCompletePump}) was prompted by features of the data that we were not initially able to explain.  Moreover, our ability to modify the parameters of the model has allowed us to identify limitations of the current source design and to explore new designs which overcome these shortcomings. We believe that the framework presented here will move forward the study and design of twin-beam sources in the high gain regime.

\section*{Acknowledgements}
G.T. thanks Merton College, Oxford, for its support.
M.D.V. thanks the Engineering and Physical Sciences Research Council
for funding through grant EP/K034480/1 (BLOQS). N.Q. and J.E.S. thank the National Science and Engineering Research
Council of Canada.

\bibliographystyle{unsrt}

\bibliography{bibliography}

\appendix
%dummy comment inserted by tex2lyx to ensure that this paragraph is not empty

\section{Quantum and classical input-output relations}

\label{appendix:classical_quantum_connection} We can exactly find the transformation that describes the Heisenberg picture evolution of the creation/annihilation operators by looking at how it transforms the classical coherent amplitudes of a seed field. A broadband coherent seed in the signal (idler) mode can be described as a displaced vacuum, which in the Heisenberg picture can represented by splitting a creation operation in terms of its mean value $\alpha_{\rm s,i}$ and its fluctuations $\delta a_{\rm s,i}$
\begin{equation}
a_{\text{s(i)}}^{\text{(in)}}(\omega)=\alpha_{\text{s(i)}}(\omega)+ \delta a_{\text{s(i)}}(\omega).
\end{equation}

The seed field is evolved following equation \ref{transferFunctions}:
\begin{widetext}
\begin{equation}
%\begin{split}
a_{\mathrm{s}}^{(\mathrm{out})}(\omega)= \int\!d\omega'\,U^{\mathrm{s},\mathrm{s}}(\omega,\omega')\,[\delta a_{\mathrm{s}}(\omega')+\alpha_{s}(\omega')]+ \int\!d\omega'\,U^{\mathrm{s},\mathrm{i}}(\omega,\omega')\,[\delta a_{\mathrm{i}}^{\dagger}(\omega')+\alpha_{i}(\omega')]%;\\
%a_{\mathrm{i}}^{(\mathrm{out})}(\omega)&=  \int\!d\omega'\,U^{\mathrm{i},\mathrm{i}}(\omega,\omega')\,[a_{\mathrm{i}}^{(\mathrm{in})}(\omega')+\alpha_{i}(\omega')]+ \int\!d\omega'\,U^{\mathrm{i},\mathrm{s}}(\omega,\omega')\,[a_{\mathrm{s}}^{\dagger(\mathrm{in})}(\omega')+\alpha_{s}(\omega')],
%\end{split}
\end{equation}
A spectrally resolved power measurement of the signal fields yields:
\begin{equation}
\begin{split}P_{s}(\omega)  \propto& \langle[a_{s}^{\text{(out)}}(\omega)]^{\dagger}a_{s}^{\text{(out)}}(\omega)\rangle  =\left|\int\!d\omega d\omega'\,U^{\text{s,s}}(\omega,\omega')\alpha_{\text{s}}(\omega')\right|^{2} +\left|\int\!d\omega d\omega'\,U^{\text{s,i}}(\omega,\omega')\alpha_{\text{i}}(\omega')\right|^{2} +\left|\int\!d\omega d\omega'\,U^{\text{s,i}}(\omega,\omega')\right|^{2},
\end{split}
\end{equation}
and a similar equation for the idler power.
%\begin{equation}
%\begin{split}P_{i}(\omega)  \propto& \langle[a_{i}^{\text{(out)}}(\omega)]^{\dagger}a_{i}^{\text{(out)}}(\omega)\rangle\\
% & =\left|\int\!d\omega d\omega'\,U^{\text{i,i}}(\omega,\omega')\alpha_{\text{i}}(\omega')\right|^{2}\\
% &\quad  +\left|\int\!d\omega d\omega'\,U^{\text{i,s}}(\omega,\omega')\alpha_{\text{s}}(\omega')\right|^{2}\\
% & \quad +\left|\int\!d\omega d\omega'\,U^{\text{i,s}}(\omega,\omega')\right|^{2}
%\end{split}
%\end{equation}
We identify three contributions to the PSD: the first, involving
the self-mode TFs, represents the seed power and cascaded DFG.
The second, involving the cross-mode TF,
represents stimulated emission (and higher order cascaded DFG terms). The third term represents
spontaneous emission. This last term is negligible for strong
enough seed fields.

\newcommand{\ii}{i}
\section{Second order perturbative expansion of PDC}
 \label{perturbative_expansion}
Here we perturbatively solve the PDC equations of motion (\ref{diffEqCompleteSignal},\ref{diffEqCompleteIdler}) up to second order in the PDC gain. We show that, while the term linear in $\gamma_\text{PDC}$ yields the usual joint spectral amplitude, the second order term yields a broadband pedestal in the same-mode transfer function. This illustrates our simple physical picture: the same mode emission is caused by a cascaded process where a broadband signal (idler) photon stimulated by the narrowband idler (signal) seed subsequently stimulates broadband
emission in the idler (signal) mode.
Our fundamental integral equations are

\begin{align} a_{\text{s}}(z,\omega)&=a_{\text{s}}(z_0,\omega)e^{i\Delta k_{\rm s}(\omega)(z-z_0)} \label{eq:int1}+\frac{i\gamma_{\text{PDC}}}{\sqrt{2\pi}}\int_{z_0}^{z}dz' g(z') e^{i\Delta k_{\rm s}(\omega)(z-z')}\int d\omega' \beta_{\text{p}}(z',\omega+\omega')a_{\text{i}}^{\dagger}(z',\omega'),  \\
a_{\text{i}}(z,\omega)&=a_{\text{i}}(z_0,\omega)e^{i\Delta k_{\rm i}(\omega)(z-z_0)}+\frac{i\gamma_{\text{PDC}}}{\sqrt{2\pi}}\int_{z_0}^{z}dz'g(z') e^{i\Delta k_{\rm i}(\omega)(z-z')}\int d\omega' \beta_{\text{p}}(z',\omega+\omega')a_{\text{s}}^{\dagger}(z',\omega').
\end{align}
The equations above are clearly formal solutions of equations  (\ref{diffEqCompleteSignal},\ref{diffEqCompleteIdler}). Taking the adjoints and
substituting into (\ref{eq:int1}) we have

\begin{align}
& a_{\text{s}}(z,\omega)=a_{\text{s}}(z_0,\omega)e^{i\Delta k_{\rm s}(\omega)(z-z_0)}+\frac{i\gamma_{\text{PDC}}}{\sqrt{2\pi}}\int_{z_0}^{z}dz' g(z') e^{i\Delta k_{\rm s}(\omega)(z-z')}\int d\omega' \beta_{\text{p}}(z',\omega+\omega')e^{-i\Delta k_{\rm i}(\omega')(z'-z_0)}a_{\text{i}}^{\dagger}(z_0,\omega')\\
& +\frac{\gamma_{\text{PDC}}^{2}}{2\pi}\int_{z_0}^{z}dz' g(z')\int_{z_0}^{z'}dz'' g^*(z'') e^{i\Delta k_{\rm s}(\omega)(z-z')}\int\int d\omega'd\omega'' \beta_{\text{p}}(z',\omega+\omega')e^{-i\Delta k_{\rm i}(\omega')(z'-z'')} \beta_{\text{p}}^{*}(z'',\omega'+\omega'')a_{\text{s}}(z'',\omega''),\nonumber 
\end{align}
and a similar equal for the evolution of $a_{\rm i}(z,\omega)$ obtained by letting $\rm s \leftrightarrow \rm i$ in the last equation. In what follows we only write equations for the signal but it is understood that a similar equation follows for the idler using the rule written in the last sentence. 
%while 
%\begin{align}
%& a_{\text{i}}(z,\omega)=a_{\text{i}}(z_0,\omega)e^{i\Delta k_{\rm s}(\omega)(z-z_0)}+\frac{i\gamma_{\text{PDC}}}{\sqrt{2\pi}}\int_{z_0}^{z}dz' g(z') e^{i\Delta k_{\rm s}(\omega)(z-z')}\int d\omega' \beta_{\text{p}}(z',\omega+\omega')e^{-i\Delta k_{\rm i}(\omega')(z'-z_0)}a_{\text{s}}^{\dagger}(z_0,\omega')\\
%& +\frac{\gamma_{\text{PDC}}^{2}}{2\pi}\int_{z_0}^{z}dz' g(z')\int_{z_0}^{z'}dz'' g^*(z'') e^{i\Delta k_{\rm s}(\omega)(z-z')}\int\int d\omega'd\omega'' \beta_{\text{p}}(z',\omega+\omega')e^{-i\Delta k_{\rm i}(\omega')(z'-z'')} \beta_{\text{p}}^{*}(z'',\omega'+\omega'')a_{\text{i}}(z'',\omega'').\nonumber 
%\end{align}

These expressions are still exact. Beginning an iteration and  looking
at a final $z=z_1$ %and setting $a_{\text{s/i}}(z'',\omega'') \to e^{i \Delta k_{\text{s/i}}(\omega'')(z''-z_0)}a_{\text{s/i}}(z_0,\omega'')$ in the right-hand side 
we have
\begin{align}
 a_{\text{s}}(z_1,\omega)=&a_{\text{s}}(z_0,\omega)e^{i\Delta k_{\rm s}(\omega)(z_1-z_0)}+\frac{i\gamma_{\text{PDC}}}{\sqrt{2\pi}}\int_{z_0}^{z_1}dz' g(z') e^{i\Delta k_{\rm s}(\omega)(z_1-z')}\int d\omega' \beta_{\text{p}}(z',\omega+\omega')e^{-i\Delta k_{\rm i}(\omega')(z'-z_0)}a_{\text{i}}^{\dagger}(z_0,\omega')\nonumber \\
& +\frac{\gamma_{\text{PDC}}^{2}}{2\pi}\int_{z_0}^{z_1}dz' g(z')\int_{z_0}^{z'}dz'' g^*(z'') e^{i\Delta k_{\rm s}(\omega)(z_1-z')}  \nonumber  \\
&\quad \times \int\int d\omega'd\omega'' \beta_{\text{p}}(z',\omega+\omega')e^{-i\Delta k_{\rm i}(\omega')(z'-z'')} \beta_{\text{p}}^{*}(z'',\omega'+\omega'') e^{i \Delta k_{\text{s}}(\omega'')(z''-z_0)} a_{\text{s}}(z_0,\omega'').\label{eq:sexact}
\end{align}
%and 
%\begin{align}
%a_{\text{i}}(z_1,\omega)=&a_{\text{i}}(z_0,\omega)e^{i\Delta k_{\rm s}(\omega)(z_1-z_0)}+\frac{i\gamma_{\text{PDC}}}{\sqrt{2\pi}}\int_{z_0}^{z_1}dz' g(z') e^{i\Delta k_{\rm s}(\omega)(z_1-z')}\int d\omega' \beta_{\text{p}}(z',\omega+\omega')e^{-i\Delta k_{\rm i}(\omega')(z'-z_0)}a_{\text{s}}^{\dagger}(z_0,\omega')\nonumber \\
%& +\frac{\gamma_{\text{PDC}}^{2}}{2\pi}\int_{z_0}^{z_1}dz' g(z')\int_{z_0}^{z'}dz'' g^*(z'') e^{i\Delta k_{\rm s}(\omega)(z_1-z')}  \nonumber  \\
%&\quad \times \int\int d\omega'd\omega'' \beta_{\text{p}}(z',\omega+\omega')e^{-i\Delta k_{\rm i}(\omega')(z'-z'')} \beta_{\text{p}}^{*}(z'',\omega'+\omega'') e^{i \Delta k_{\text{i}}(\omega'')(z''-z_0)} a_{\text{i}}(z_0,\omega''),\label{eq:sexact1}
%\end{align}
Then writing %$z_0 = \ell_{\text{min}}, z_1 = \ell_{\text{max}}$ and 
\begin{align*}
 a^{\text{(out)}}_{\text{s}}(\omega)&=\int d\omega''U^{s,s}(\omega,\omega'')a^{\text{(in)}}_{\text{s}}(\omega'')+\int d\omega'U^{s,i}(\omega,\omega')a_{\text{i}}^{\dagger (\text{in})}(\omega'),\\
 a^{\text{(out)}}_{\text{i}}(\omega)&=\int d\omega''U^{i,i}(\omega,\omega'')a^{\text{(in)}}_{\text{i}}(\omega'')+\int d\omega'U^{i,s}(\omega,\omega')a_{\text{s}}^{\dagger \text{(in)}}(\omega'),\\
a_{\rm x}^{\text{(in)}}(\omega)&= e^{-i \Delta k_{\rm x}(\omega) z_0} a_{\rm x}(z_0,\omega), \quad a_{\rm x}^{\text{(out)}}(\omega)= e^{-i \Delta k_{\rm x}(\omega) z_0} a_{\rm x}(z_1,\omega)
\end{align*}
we have
\begin{align}
& U^{\rm s,s}(\omega,\omega'')=\delta(\omega-\omega'')\\
& +\frac{\gamma_{\text{PDC}}^{2}}{2\pi}\int_{z_0}^{z_1}dz' g(z') \int_{z_0}^{z'}dz'' g^*(z'') e^{-i\Delta k_{\rm s}(\omega)z'}\int d\omega' \beta_{\text{p}}(z',\omega+\omega')e^{-i\Delta k_{\rm i}(\omega')(z'-z'')} \beta_{\text{p}}^{*}(z'',\omega''+\omega')e^{i\Delta k_{\rm s}(\omega'')z''}, \nonumber \\
& U^{\rm s,i}(\omega,\omega')=\frac{i\gamma_{\text{PDC}}}{\sqrt{2\pi}}\int_{z_0}^{z_1} dz'e^{-i\Delta k_{\rm s}(\omega)z'}g(z') \beta_{\text{p}}(z',\omega+\omega')e^{-i\Delta k_{\rm i}(\omega')z'}.
%& U^{\rm i,i}(\omega,\omega'')=\delta(\omega-\omega')  \\
%& +\frac{\gamma_{\text{PDC}}^{2}}{2\pi}\int_{\ell_{\text{min}}}^{\ell_{\text{max}}}dz' g(z') \int_{\ell_{\text{min}}}^{z'}dz'' g^*(z'') e^{-i\Delta k_{\rm i}(\omega)z'}\int d\omega'' \beta_{\text{p}}(z',\omega+\omega')e^{-i\Delta k_{\rm s}(\omega')(z'-z'')} \beta_{\text{p}}^{*}(z'',\omega''+\omega')e^{i\Delta k_{\rm i}(\omega'')z''}, \nonumber\\
%& U^{\rm i,s}(\omega,\omega')=\frac{i\gamma_{\text{PDC}}}{\sqrt{2\pi}}\int_{\ell_{\text{min}}}^{\ell_{\text{max}}}dz'e^{-i\Delta k_{\rm i}(\omega)z'}g(z') \beta_{\text{p}}(z',\omega+\omega')e^{-i\Delta k_{\rm s}(\omega')z'}.
\end{align}
\end{widetext} 
The first-order perturbative result in $U^{\rm s,i}$  in the last set of equations describes how a field in the idler (signal) mode stimulates the generation
of a field in the signal (idler) mode, and is often employed in the context
of photo-pair generation, where the first order approximation of $U^{\rm s,i (i,s)}(\omega,\omega')$ is the joint spectrum of SPDC photon pairs.
This can be written in a more elegant form by introducing the phase-matching function (PMF) 
\begin{align}
\Phi(\Delta k(\omega,\omega')) &= \int_{z_0}^{z_1} \frac{dz}{\sqrt{2\pi}} g(z) e^{- i\Delta k(\omega,\omega') z}, \\
\Delta k(\omega,\omega')  &= \Delta k_{\rm s}(\omega)+ \Delta k_{\rm i}(\omega')\\
&=(\Delta \beta_{\text{s}})(\omega - \bar{\omega}_{\rm s})+(\Delta \beta_{\text{i}})(\omega' - \bar{\omega}_{\rm i}), \\
\Delta \beta_{\text{x}} &= 1/v_{\text{x}} - 1/v_{\text{p}}, \label{deltabetadef}
\end{align}
and recalling that in the low gain regime the spectral content of the pump is not changed allowing us to write
\begin{align}
\beta_{\rm p}(z,\omega+\omega') = \sqrt{\frac{E_{\rm p}}{\sigma}} F\left(\frac{\omega-\bar{\omega}_{\rm s} + \omega'-\bar{\omega}_{\rm i} }{\sigma} \right),
\end{align}
where $F(x)$ is the $\mathcal{L}^2$ normalized pump shape (i.e. $\int dx |F(x)|^2=1$) and we used the fact that $\bar{\omega}_{\rm p } = \bar{\omega}_{\rm s }+\bar{\omega}_{\rm i }$. With these definitions we finally write
\begin{align}
U^{\rm s,i}(\omega,\omega') =& i \gamma_{\text{PDC}}\sqrt{\frac{E_{\rm p}}{\sigma}}  F\left(\frac{\omega + \omega'-\bar{\omega}_{\rm p} }{\sigma} \right)  \Phi(\Delta k(\omega,\omega')).
\end{align}
Note that if, for example,  $g(z)$ is a top hat function of length $L$ centered at the origin then
\begin{align}
\Phi(\Delta k(\omega,\omega')) = \frac{L}{ \sqrt{2 \pi}} \text{sinc}\left( \Delta k(\omega,\omega') L/2 \right).
\end{align}

The mean number of spontaneously generated signal (idler) photons per pump pulse is thus
\begin{align}
\langle n_{\rm s(i)} \rangle &= \int d\omega \, \langle a_{\rm s}^\dagger(\omega) a_{\rm s}(\omega) \rangle = \int d\omega \, d\omega' |U^{\rm s,i}(\omega, \omega')|^2  \nonumber \\
&= \frac{\gamma_{\rm PDC}^2 E_{\rm p} \ell }{\left|\frac{1}{v_{\rm s}} - \frac{1}{v_{\rm i}}  \right|},
\end{align}
where $\ell = \int_{\ell_{\text{min}}}^{\ell_{\text{max}}} dz |g(z)|^2$ is the effective length of the crystal. Note that a $g(z)$ with a top hat shape of length $L$ will have precisely that $\ell = L$.

In a stimulated experiment where the idler (signal) is seeded, broadband DFG will be generated in the conjugate mode (Appendix \ref{appendix:classical_quantum_connection}) with a total intensity $I_{\rm s(i)} \propto \int d\omega \, d\omega'\, |U^{\rm s,i (i,s)}(\omega, \omega')|^2 = \langle n_{\rm s(i)}\rangle$.  

The second order term which appears in $U^{\rm s,s}(\omega,\omega')$ and $U^{\rm i,i}(\omega,\omega')$ generates a broadband contribution in the seeded mode with amplitude proportional to the square of the PDC gain. 

Under certain circumstances it is possible to evaluate this term analytically. For this we will assume a Gaussian nonlinearity profile and pump profile
\begin{align}
g(z) = \frac{1}{\sqrt[4]{\pi \gamma/2}} e^{-\tfrac{z^2}{\gamma \ell^2}}, \quad F(x) = \frac{1}{\sqrt[4]{\pi}} e^{-\tfrac{x^2}{2}},
\end{align}
where $\gamma \approx 0.193$ is chosen so that the PMF associated with this nonlinearity profile has the same full width at half maximum of the one associated with a top hat profile. Under these approximations, letting $z_0 \to -\infty$ and $z_1 \to +\infty$ one can write the intermode transfer function as a double Gaussian
\begin{align}
U^{\rm s,i}(\omega,\omega') &= i \gamma_{\text{PDC}} \ell \sqrt{\frac{E_{\rm p}}{\pi \sigma} \sqrt{\gamma/2}} \exp\left(- \mathbf{v} \mathbf{M} \mathbf{v}^T \right), \\
\mathbf{v} &= (\omega-\bar{\omega}_{\rm s}, \omega'-\bar{\omega}_{\rm i}) , \\
\mathbf{M} & =  \begin{pmatrix}
\mu_{\rm s}^2 & \mu^2 \\
\mu^2 & \mu_{\rm i}^2
\end{pmatrix}, \\
\mu_{\rm s,i}^2 &= \frac{1}{4} \ell^2 \gamma  \Delta \beta _{\rm s,i}^2+\frac{1}{2 \sigma^2}, \\
\mu^2 &= \frac{1}{4} \gamma  \Delta \beta _{\rm i}
\Delta \beta _{\rm s} \ell ^2+\frac{1}{2 \sigma^2}.
%\tau  &= \frac{1}{\sqrt{2} \sigma}
\end{align}
Similarly, one can write a simple analytical expression for the second-order intramode transfer function
\begin{subequations}\label{eq:USSb}
\begin{align}
U^{\rm s,s}(\omega,\omega'') =&  \delta(\omega-\omega'') + U^{\rm s, s}_{\text{b}}(\omega,\omega'') ,  \\
U^{\rm s ,s}_{\text{b}}(\omega,\omega'') =& \frac{ \gamma_{\rm PDC}^2 E_{\rm p}
	\ell^2 \sqrt{\gamma } }{\sqrt{\pi } \sigma  \mu
	_{\rm i}} e^{-x_-^2-x_+^2} \left(1+i \ 
	\text{erfi}\left(x_+\right)\right),\\
	x_+ &= \frac{ \frac{\sqrt{\gamma} \ell}{2} \left(\frac{1}{v_{\rm s}} - \frac{1}{v_{\rm i}} \right) }{2 \sigma \mu_{\rm i}}  
	(\omega+ \omega'' - 2 \bar{\omega}_{\text{s}}), \\
	x_- &= \frac{\mu_{\rm s} }{\sqrt{2}} (\omega- \omega''),
	\end{align}	
\end{subequations}
and $\text{erfi}(x) = \text{erf}(i x)/i$ is the imaginary error function, which is an odd function of its argument. 
Because of this the phase of $U^{\rm s, s}_{\text{b}}(\omega,\omega'')$ has a jump when crossing the line $\omega+ \omega'' - 2 \bar{\omega}_{\text{s}}=0$. This will become a useful observation when we compare the form of the transfer function of the cascaded (second order) $\chi^{(2)}$ process, which as just mentioned as a phase jump, and the pure $\chi^{(3)}$ process which does not.
Finally, note that an analogous expression for the intramode transfer function $U^{\rm i, i }_{\text{b}}(\omega,\omega'')$ can be obtained by letting $\rm{i} \leftrightarrow \rm{s}$ in Eq.~\eqref{eq:USSb}.
\section{Experimental setup details}
\label{appendix_experimental_details}

The nonlinear crystal in our experiment is an 8mm ppKTP chip manufactured
by ADVR, made from a z-cut wafer, with a set of waveguides written
using a proton exchange process. The waveguides are roughly $4\mu m\times4\mu m$
in cross section. The chip output is AR coated for 1550 nm. We pump
the source with a pulsed Ti:Sapphire laser (Coherent Chameleon) with
central wavelength 783 nm and 8nm bandwidth, which we filter using
two bandpass thin-film filters (Semrock LL01-785-12.5) to a bandwidth
of 1.9 nm. The periodic poling has a spatial period of $104\mu$m,
which leads to quasi-phase matching for signal and idler around 1560
nm. Using an ordinarily polarized pump (H polarization), we obtain
a signal (H polarization) and idler (V polarization) beams with respective
central wavelengths of 1563 nm and 1569 nm. 

\begin{figure}[h]
	\includegraphics[width=0.45\textwidth]{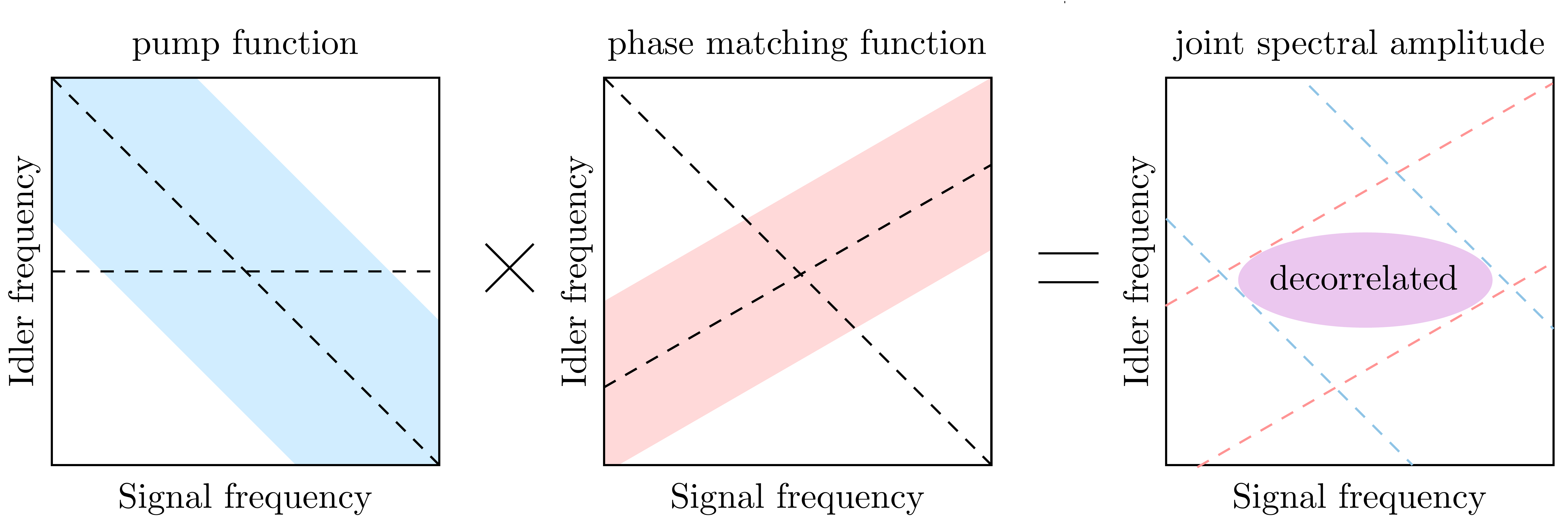}
	\caption{The product of a pump function (energy conservation) and a phase-matching
		function (momentum conservation) can yield a separable joint spectrum }
	\label{separable_phase_matching} 
\end{figure}

The phase-matching of the type-II PDC process has been engineered
to obtain signal and idler fields at telecom wavelengths with an almost
separable joint spectrum. In the weak squeezing regime, where we may approximate that only single photon pairs are produced, the joint spectrum of a PDC source can
be expressed as a product of two factors: a pump function and a phase
matching function \cite{Grice97}. As illustrated in Figure \ref{separable_phase_matching},
the pump function, related to energy conservation, introduces
anti-correlation of the frequencies of photons generated in the PDC
process. The nature of the phase matching function, related to
momentum conservation, depends on the dispersion relation of the guided
modes. In a medium with normal dispersion, both signal and idler commonly
have higher group velocity than the pump, resulting in frequency anti-correlations
\cite{MosleyThesis}. In the waveguided ppKTP, due to the birefringence
of the material the group velocity of a pump field at 780 nm in the
ordinary polarization falls in between the group velocities of the
signal and idler fields centered around 1560 nm in orthogonal polarizations.
This results in a phase-matching function with positive frequency
correlations \cite{Eckstein11} . The combined effect of these two
factors with the correct balance of pump and phase matching bandwidths,
as illustrated in Figure \ref{separable_phase_matching}, can produce
an almost separable joint spectrum. 

\begin{figure}[h]
	\includegraphics[width=0.35\textwidth]{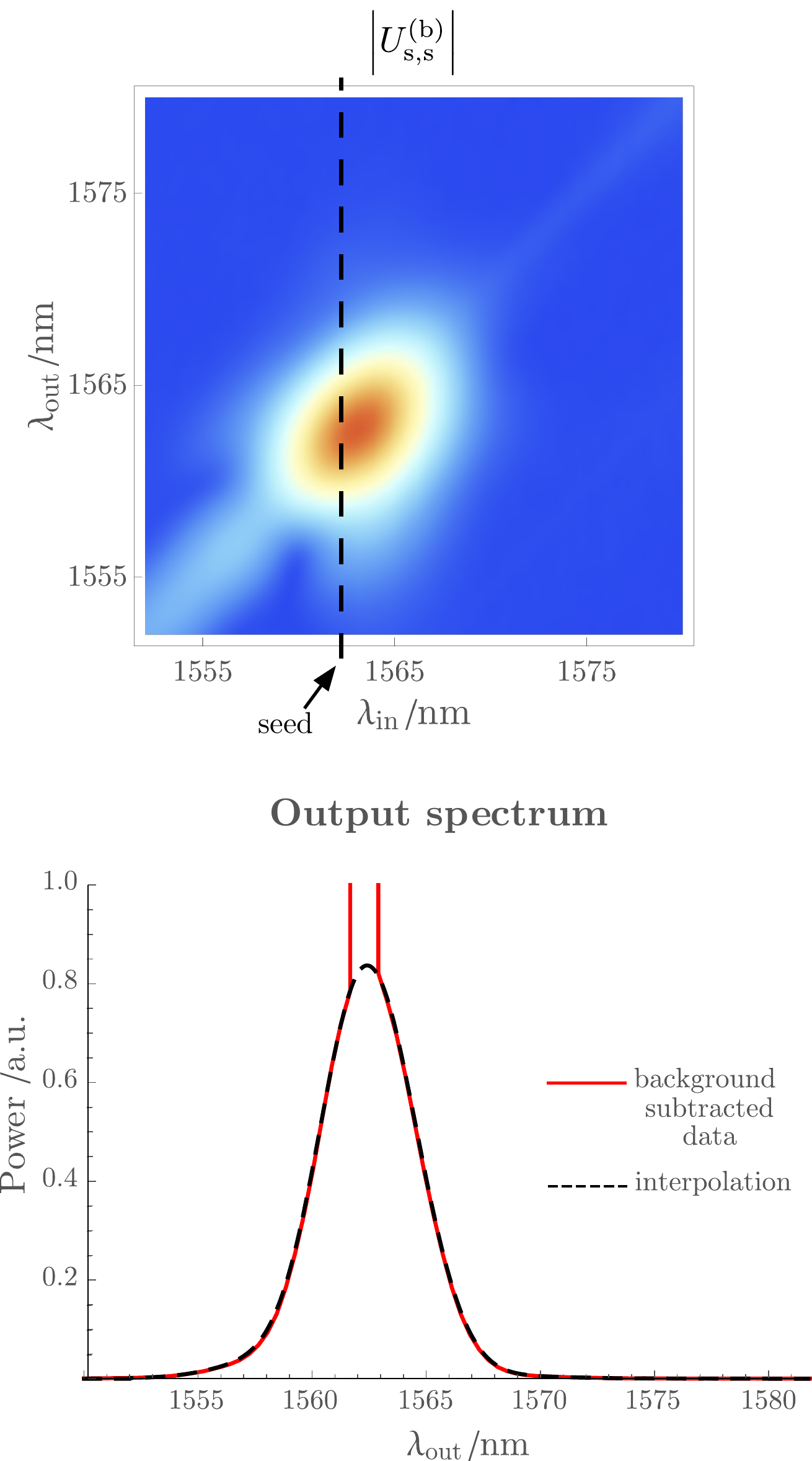}
	\caption{a). Reconstruction of the stimulated emission spectrum as a function
		of seeding wavelength, measured in the seed polarization; color indicates
		the absolute value of the field (square root of the measured spectral
		power density) b). Red -- One "slice" of the background subtracted data
		, showing a noisy measurement around the seed wavelength, where
		the OSA detector is blinded by the seed; black -- reconstructed spectrum obtained
		by interpolating over the noisy region .}
	\label{cutUss}
\end{figure}

To measure the direct and cascaded DFG signals we seed one of the
downconverted fields with a continuous wave (CW) laser and measure
the generated stimulated emission using an optical spectrum analyzer
(OSA). For the seed field we use a Yenista Tunics T100S-HP CW fiber
laser with wavelength tunable between 1500nm and 1680nm. The laser
has a very high spontaneous noise suppression,  over 100
dB over a bandwidth of 1 nm. After polarization filtering using birefringent
waveplates and a Glan-Taylor polarizer, we record spectra of the generated
light using a Yokogawa AQ6370D optical spectrum analyzer (OSA). The
OSA has a high dynamic range of 78 dB over 1nm, allowing us to distinguish
the weak broadband cascaded DFG signal ($<$1nW) from the narrowband
seed ($\approx600\mu$W). While the linewidth of the seed laser is
many orders of magnitude smaller than the broadband cascaded DFG emission,
the finite resolution of the OSA (0.2 nm) broadens the width of the
narrowband component of the spectra, hindering the retrieval of the
broadband pedestal. To remove the CW seed from the data we first subtract
a background trace from it, where we send the seed beam but not the
pump through the crystal. Fluctuations of the seed power between the
moment when the data and the background are measured lead to an imperfect
extinction of the narrowband component. We eliminate this CW remnant
by masking the frequency range occupied by it and interpolating the
remaining spectrum, as illustrated in Figure \ref{cutUss}.

\section{Identifying the EOM parameters}
\label{identifying_EOM_parameters}

In this Appendix, we describe in detail how we identify the physical
parameters that enter in our theoretical description (\ref{diffEqCompleteSignal},\ref{diffEqCompleteIdler},\ref{diffEqCompletePump})
of the SPDC. We describe how these parameters can be extracted from
the TF data mostly in the small gain limit,  complemented with some
additional measurements.

\subsection{Pump spectral intensity}
\label{parameters_pump_intensity} 

The pump spectral amplitude at position $z$ in the crystal, $\beta_{\text{p}}(z,\omega)$,
is a complex-valued function. In our experiment,
we measure the absolute value of $\beta_{\text{p}}(z=0,\omega)$
in two ways (see figure \ref{pumpPhase}): First, we perform a direct measurement of the pump spectrum at the chip input
using an optical spectrum analyzer (OSA), which provides the squared
modulus  of the pump spectral amplitude. Second, we confirm this result by extracting the pump spectral density from low gain cross-mode TF measurements.

\subsection{Pump spectral phase}
\label{parameters_pump_phase} 
We used an APE LX-Spider (wavelength range
between 750 and 900 nm), to characterize the spectral phase of the
Ti:Sapphire pump laser, but could not extend that measurement to the
filtered pump (1.9 nm bandwidth) because of the limit in the spectral
resolution of the apparatus. We characterized the spectral phase added
by the filters using a spectral self-interference measurement: the
filtered pump is interfered with a sample of the same, unfiltered field,
thus revealing their differential spectral phase. The total spectral phase of the pump is obtained by adding the spectral phase of the reference beam, measured using SPIDER, and the phase added by the optical filters, measured through spectral interference.

Let us briefly detail the spectral interference process by which we measure the phase added by the optical filters. The complex spectral amplitude corresponding to the sum of the unfiltered reference, $\beta_\text{ref}$, and filtered pump field, $\beta_\text{p}(\omega)$, is
\begin{equation}
\beta_\mathrm{sum}(\omega)= \beta_\mathrm{ref} + |\beta_\mathrm{p}(\omega)| \, \exp\Big(i(\phi(\omega)+ \Delta t \, \omega)\Big)  ,
\end{equation}
where we are omitting the spatial label in the field amplitudes, implicitly assuming a position before the chip input.  We assume that the unfiltered reference is much broader than the filtered pump,
such that the former can be treated as a field with a constant spectral density. Without loss of generality, we also take the spectral phase as constant. The phase of the filtered pump is the sum of $\phi(\omega)$, the phase added in the filtering process, and $\Delta t \, \omega$, representing a temporal delay with respect to the reference field.

The measured PSD is:

\begin{align}
&P(\omega) \propto |\beta_\mathrm{sum}(\omega)|^{2} \\
&=|\beta_\mathrm{ref}|^{2} + |\beta_\mathrm{p}(\omega)|^2  + |\beta_\mathrm{ref}| | \beta_\mathrm{p}(\omega)| \cos\big(\phi(\omega) - \Delta t \, \omega\big),  \nonumber 
\end{align}
where we recognize that the last term describes spectral fringes with a period of approximately $1/\Delta t$, modulated by the spectral phase added by the optical filters. We assume that the delay between the interfering beams is large, so
that the interference fringes that it causes are fast compared with
the spectral interference due to the filters' spectral phase. This modulation can be retrieved by keeping only positive frequencies above the bandwidth of the base-band component of $I(\omega)$, which corresponds to the complex signal
\begin{equation}
P_\mathrm{HF} \propto |\beta_\mathrm{ref}| |\beta_\mathrm{p}(\omega)| \exp\Big(\Delta t \, \omega + \phi(\omega)\Big).
\end{equation}
The argument of which corresponds to the spectral phase added by the filters, modulo a linear component. 

\begin{figure}[h]
	\includegraphics[scale=0.425]{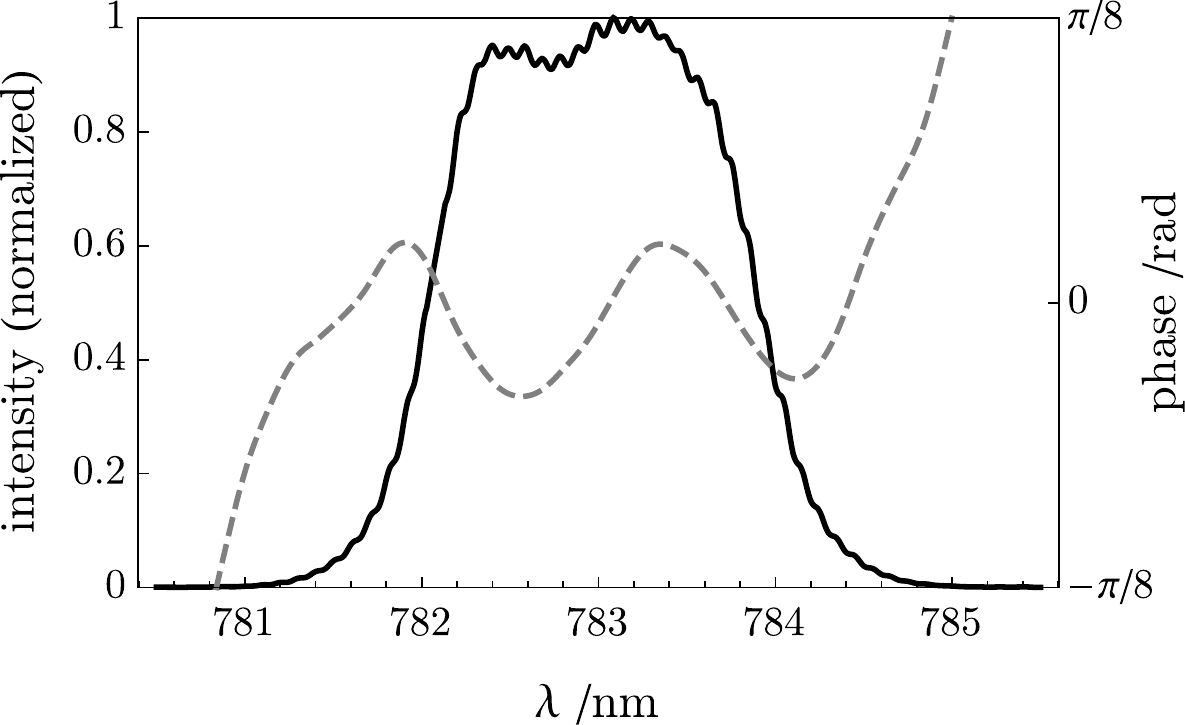}
	\caption{ Magnitude (solid line) and phase (dashed line) of the filtered pump spectral amplitude. The magnitude shown here corresponds to that measured using the OSA, and agrees very well with the pump function of a low gain JSI. The phase shown here corresponds to the addition of the phase of the unfiltered pump measured using SPIDER, and the differential phase of the filtered pump measured through spectral interferometry.}
	\label{pumpPhase} 
\end{figure}

Our SPIDER measurement reveals that the chirp
introduced in the unfiltered pump by the power-distribution optics
is small, reaching less than 0.1 rad over the filtered pump bandwidth. The
spectral phase added by the dielectric filters is the main contribution
to the total spectral phase. This measurement agrees with the specifications
provided by the supplier of the filters (Semrock). In figure \ref{pumpPhase} we show the measured magnitude and phase of the pump spectral amplitude. 

\subsection{Group velocity mismatch}

\label{parameters_group_velocity_mismatch}

In our model   the group velocities of the pump, signal, and idler
fields  determine the linear propagation of the fields in the crystal.
We need the two group velocity mismatch
parameters, $\Delta\beta_{\text{s(i)}}=1/v_{\text{s(i)}}-1/v_{\text{p}}$.  The phase matching (PM) function (see Appendix \ref{perturbative_expansion}) of a low gain joint spectral amplitude has the simple form 
\begin{equation}
\Phi_\text{PM}(\Omega_{ \rm s}, \Omega_{\rm i}) \propto \text{sinc}\left(\frac{L}{2} (\Delta \beta_{\rm s} \Omega_{\rm s} + \Delta \beta_{\rm i} \Omega_{\rm i})\right),
\end{equation}
where $L$ is the crystal length and  $\Omega_{\rm s(i)} = \omega_{\rm s(i)} - \bar{\omega}_{\rm s(i)}$ are the detunings from the central phase-matching frequencies. Perfect phase-matching occurs for frequencies $\Omega_{\rm i} = \Omega_{\rm s} \tan(\theta_\text{PM})$ , with  $\tan(\theta_\text{PM})= -\Delta \beta_\text{\rm s}/\Delta \beta_\text{\rm i}$. Rotating to a frequency frame by the phase matching angle, $\theta_\text{PM}$,
\begin{eqnarray}
&\Omega_{||} = \cos(\theta_\text{PM}) \Omega_{\rm s} +  \sin(\theta_\text{PM}) \Omega_{\rm i}, \\ &\Omega_{\perp} = \cos(\theta_\text{PM}) \Omega_{\rm i} -  \sin(\theta_\text{PM}) \Omega_{\rm s} , 
\end{eqnarray}
the phase-matching function depends only on the perpendicular component:
\begin{equation}
\Phi_\text{PM}(\Omega_{\perp}) \propto \text{sinc}\left(\frac{1}{2} \Omega_{\perp} \frac{\Delta\mu}{\cos(\theta_\text{PM}) + \sin(\theta_\text{PM})}\right),
\label{phase_matching_profile}
\end{equation}
where we have introduced the quantity $\Delta\mu= L(\Delta \beta_\text{s}-\Delta \beta_\text{i}) = L(1/v_\text{s}-1/v_\text{i})$, which corresponds to the group delay between the signal and idler over the crystal length. This parametrization makes it clear that, in the symmetric group velocity matching case, the phase-matching bandwidth is inversely proportional to the walk-off between the signal and idler pulses in the crystal. We can write the group velocity mismatch parameters, $\Delta \beta_{\rm s(i)}$, in terms of $\Delta \mu$ and $\theta_{\text{PM}}$ as
\begin{subequations}
	\begin{align}
\Delta \beta_\text{s} &=  \frac{\Delta\mu}{L} \frac{\tan(\theta_\text{PM})}{\tan(\theta_\text{PM})+1}, \label{GVM_from_PM_angle_and_bandwidth1} \\
 \Delta \beta_\text{i} &= - \frac{\Delta\mu}{L} \frac{1}{\tan(\theta_\text{PM})+1}.
\label{GVM_from_PM_angle_and_bandwidth2}
\end{align}
\end{subequations}

\begin{figure}[h]
	\includegraphics[scale=0.22]{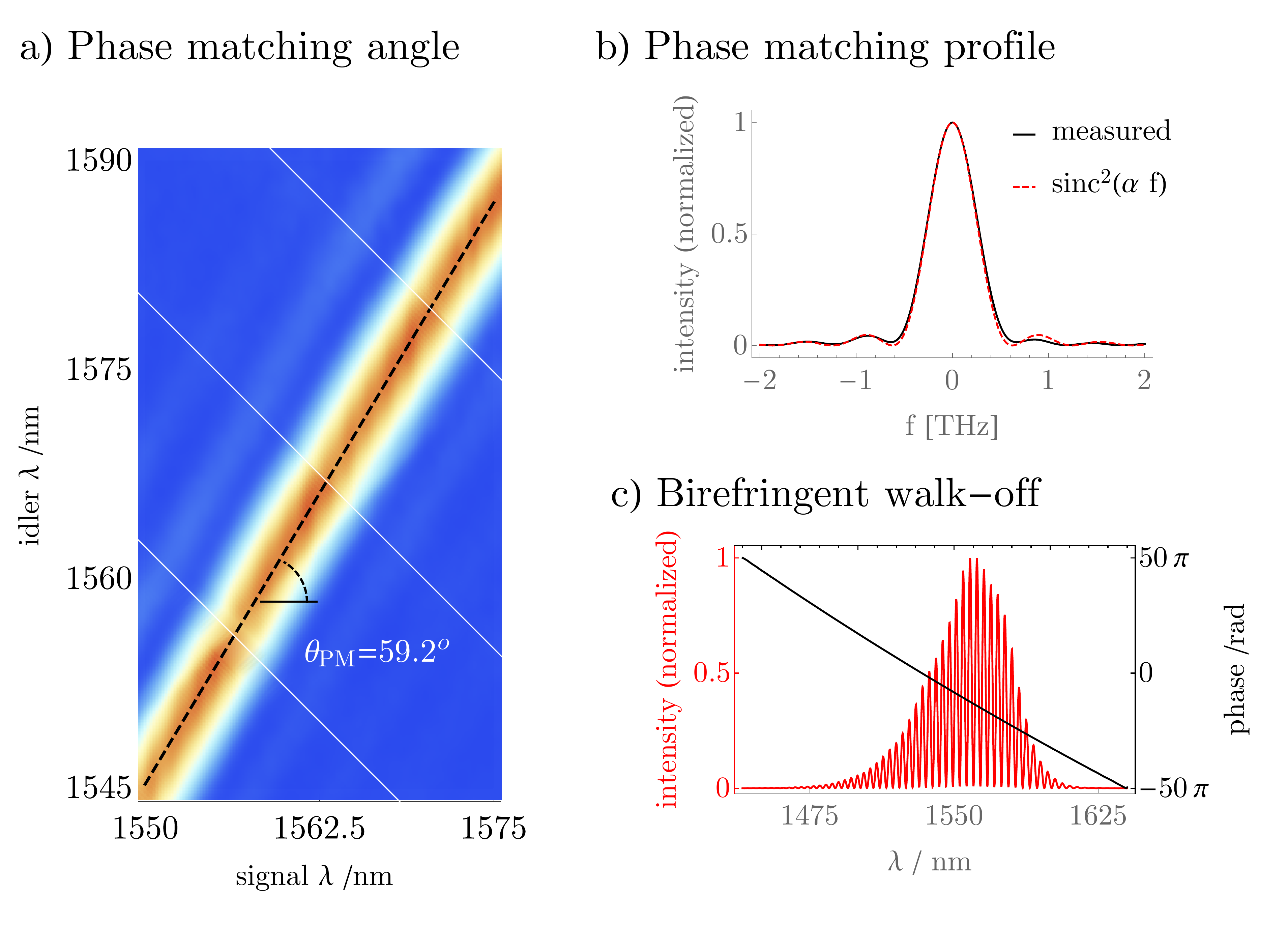}
	\caption{Group velocity mismatch characterization. a) Measurement of the phase matching angle. Four low gain JSIs with different pump central frequencies are joined to reconstruct a long section of the phase matching function. b) Inference of $\Delta \mu$ by measuring the bandwidth of the phase matching profile. The measured intensity profile (solid black line) is fitted with a $\text{sinc}^2$ function (dashed red line), and the group delay is obtained from equation (\ref{phase_matching_profile}). c) Direct measurement of $\Delta \mu$ by measuring spectral interference between two orthogonal polarizations of a broadband SLED after travelling through the crystal. The oscillating spectral intensity (red line) is transformed to the Fourier domain and a corresponding spectral phase is obtained (black line), from which the spectral oscillation period is retrieved. }
	\label{dispersion_characterization} 
\end{figure}

To obtain the phase matching angle, $\theta_{\text{PM}}$, we reconstruct
a broad section of the phase matching function by measuring the cross-mode
TF as we scan the pump power and joining the resulting measurements
(see Figure \ref{dispersion_characterization}.a). We found the phase-matching
angle $\theta_{\text{PM}}=59.2^{\circ}$.

The second parameter, $\Delta\mu$, can be obtained by measuring the
spectral interference due to walk-off between pulses in the signal and idler polarizations propagating through the source. We launched broadband light from a very broadband ($\sim 100$ nm) superluminiscent light-emitting diode (SLED) in diagonal polarization through the ppKTP wave-guide. The input polarization state can be written as $\ket{D}=\left(\ket{H}+\ket{V}\right)/\sqrt{2}$. The output polarization state is then $\left(e^{i\,\phi_{\rm H}}\ket{H}+e^{i\,\phi_{\rm V}}\ket{V}\right)/\sqrt{2}$. Up to first order in the wavelength $\lambda$, we have
\begin{equation}
\begin{split}
&\phi_{\rm V} - \phi_{\rm H} =  \\
&\omega_0\,L\left[\left(\frac{n_{\rm V}(\omega_0)-n_{\rm H}(\omega_0)}{c}+\frac{1}{v_{\rm i}}-\frac{1}{v_{\rm s}}\right)+\left(\frac{1}{v_{\rm s}}-\frac{1}{v_{\rm i}}\right)\frac{\lambda}{\lambda_0}\right] \\
&={\rm constant} + \frac{2\pi c}{\lambda_0^2}\,\Delta \mu\,\lambda,
\end{split}
\end{equation}
where $\omega_0$ is the central frequency of the signal/idler fields, $\lambda_0$ is the corresponding central wavelength in vacuum, and $n_\text{V/H}(\omega_0)$ is the refractive index experienced by the vertical/horizontal polarizations of a field at said central wavelength. Thus, when we project the output on the diagonal polarization, in the output wavelength spectrum we will observe fringes with period $ \lambda_0^2/(c \,  \Delta \mu)$. Figure \ref{dispersion_characterization}.c shows our measurement of
these spectral fringes, from which we extract a delay between signal
and idler of $\Delta\mu=2.2$ ps.

\begin{figure}[h]
	\includegraphics[scale=0.375]{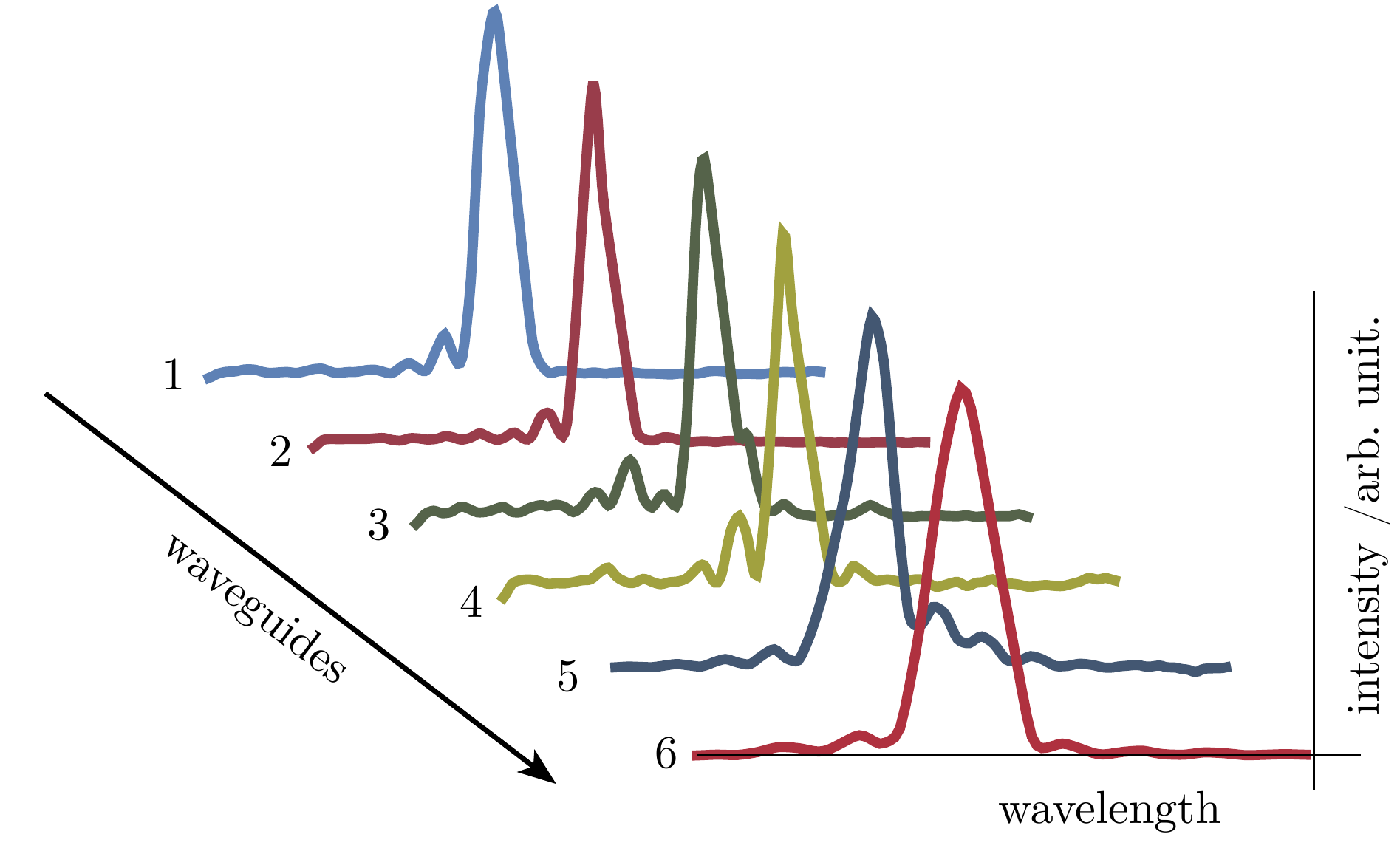}
	\caption{Phase matching profiles of different waveguides in the same chip. We observe a variability of the shape of the phase-matching profiles which can be caused by periodic poling inhomogeneity.}
	\label{inhomogeneous_phase_matching_profile} 
\end{figure}

As shown in Figure \ref{dispersion_characterization}.b, we independently infer the value of $\Delta \mu$ by measuring the phase-matching profile of a low gain JSI  and fitting it to the function predicted in Equation (\ref{phase_matching_profile}). While it is known that inhomogeneities in the periodic poling of a nonlinear
crystal lead to distortions of the phase-matching profile \cite{Fejer92,Guo14}, the good fit of the data to a ``sinc'' function allows us to assume that the periodic poling in this waveguide does not show a significant inhomogeneity. We must note that this was not the case in other waveguides of the same chip, which yielded phase-matching profiles showing important deviations from a ``sinc'' function (see Figure \ref{inhomogeneous_phase_matching_profile}). The value of the group delay that fits the bandwidth of the measured phase-matching profile is again $\Delta \mu = 2.2$ ps, confirming our previous measurement.

The values of the group velocity mismatch parameters retrieved from Equations (\ref{GVM_from_PM_angle_and_bandwidth1},\ref{GVM_from_PM_angle_and_bandwidth2}) together with the results of our measurements, and assuming a crystal of length of $L = 8$mm, are $\beta_s = 1.7$ ps/cm and $\beta_i=-1.0$ ps/cm.
\subsection{Parametric down-conversion gain}
\label{parameters_PDC_gain}

In the low gain regime, where the effect of SPM and XPM can be neglected, and assuming a homogeneous periodic poling, the PDC equations of motion (\ref{diffEqCompleteSignal}, \ref{diffEqCompleteIdler}) are defined solely by $\Delta \beta_{\rm s}, \Delta \beta_{\rm i}$ and $\beta_\text{p}(z=0, \, \omega)$, as well as the PDC coupling strength,  $\gamma_\text{PDC}$. Integrating the EOMs allows us to compute the quantity $\kappa$, as defined in equation (\ref{ratio}),
which, as we have explained in Sec \ref{extracting_TFs_from_DFG}, is independent of the seeding and detection efficiency, using $\gamma_\text{PDC}$ as a free fitting parameter. From the results in Appendix \ref{perturbative_expansion}, one can easily show that $\kappa$ is linearly proportional to $\gamma_\text{PDC}$ in the low gain regime. In the high gain regime, the relation between $\kappa$ and $\gamma_\text{PDC}$ becomes nonlinear, but it remains monotonically increasing. This monotonic dependence allows us to use $\kappa$ as a robust proxy to derive the PDC coupling strength, $\gamma_\text{PDC}$, from experimental intensity ratios.

In our experimental demonstration, we fit the predictions of our model to the experimental values of
$\kappa$ as a function of pump pulse energy, as illustrated in figure
\ref{ratiosFigure}. In order to mitigate the effect of measurement noise, we smooth the spectral distributions using a Gaussian kernel with a standard deviation of 0.35 nm before taking their maxima. The best fit parameter was $\gamma_\text{PDC} = 28 \, \text{W}^{-1/2}/\text{m}^{-1}$.

\subsection{Cross-phase modulation}
\label{parameters_XPM}
\begin{figure}[h]
	\includegraphics[scale=0.3]{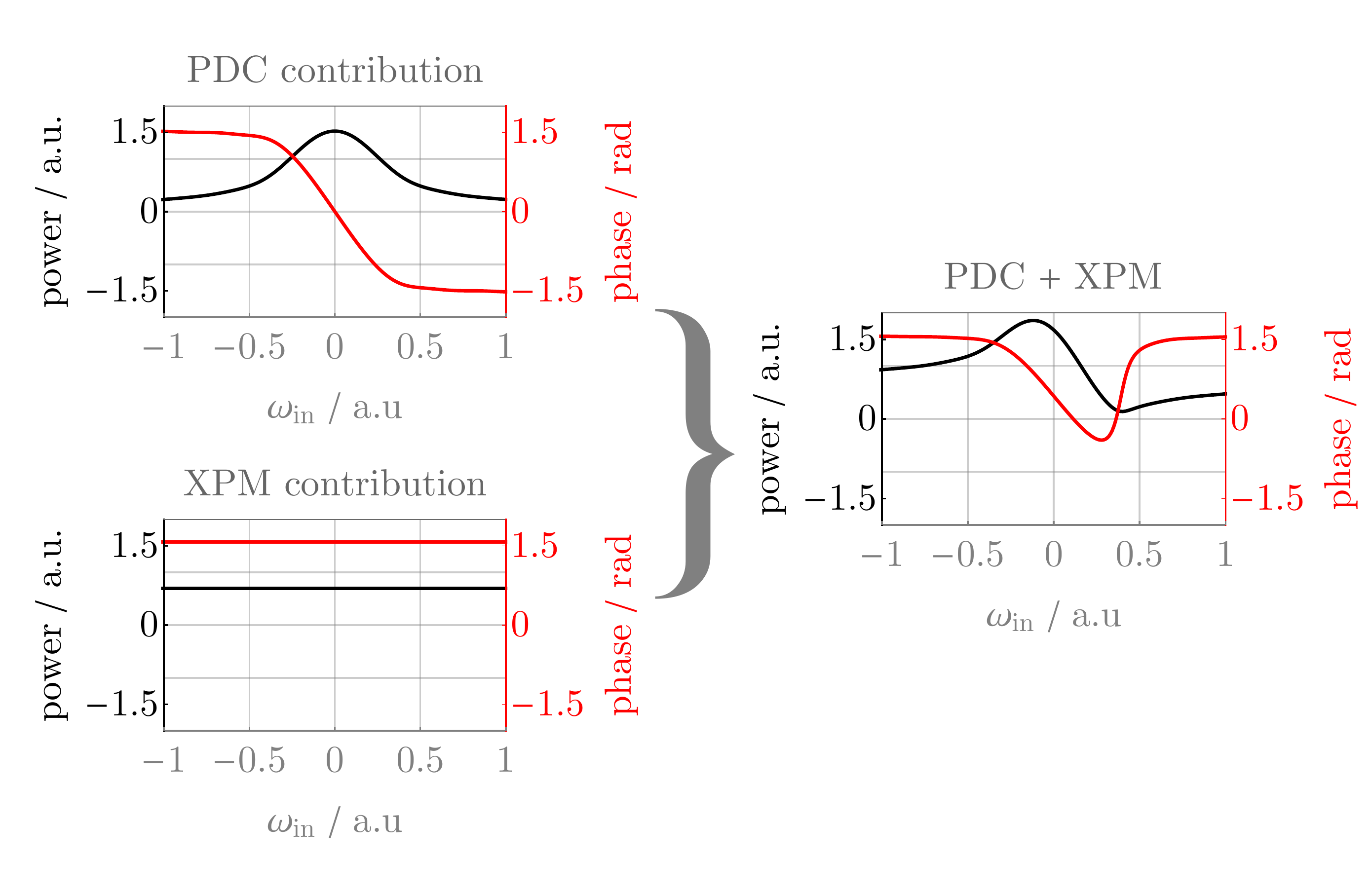}
	\caption{Asymmetric magnitude of the TF ``cut" $U^{s,s}_{(b)}(\omega, \omega)$ due to interference between PDC and XPM. Left, top -- Simulated cut with only PDC. Left, bottom -- Simulated cut with only XPM. Right -- Simulated cut with PDC and XPM.  }
	\label{simple_pdc_xpm}
\end{figure}

Cross-phase modulation of the narrowband seed by the broadband pump appears as a spectral broadening of the former. In our seeded measurements, this effect shows as a broadband pedestal surrounding the CW component of the same-mode TFs. According to our simulations of $U_{\rm b}^{\rm s,s/i,i}(\omega_{\text{out}},\omega_{\text{in}})$, the amplitude  is constant along a contour where $\omega_\text{in} + \omega_\text{out}$ is constant, and its magnitude is proportional to the power of the pump. As mentioned in previous appendices (e.g Appendix \ref{perturbative_expansion}), the same-mode TFs contain another broadband component due to cascaded PDC. Our simulations show that the cascaded PDC amplitude scales linearly with the pump power and it exhibits a $\pi$ phase jump around the central phase-matching wavelength (this is also readily see by examining the analytical result in Eq.~\eqref{eq:USSb}). Due to this non-trivial spectral phase, the XPM and PDC contributions interfere either constructively or destructively at different frequencies, resulting in an emission spectrum shaped like a Fano resonance \cite{Fano61}, as illustrated in figure \ref{simple_pdc_xpm}. The asymmetry of this spectrum allows us to accurately estimate the XPM coupling strength in relation to the PDC coupling strength. In Figure \ref{XPMfitting} we use diagonal cuts of the same-mode TFs at the lowest power (in order to avoid other nonlinear phenomena) and find the XPM interaction strength $\gamma_{\mathrm{XPM,s(i)}}$ that best fits their asymmetric ``tails". The best fit parameters were $\gamma_\text{XPM,s} = 0.16 \, \text{W}^{-1} \text{m}^{-1}$ and $\gamma_\text{XPM,i} = 0.06 \,  \text{W}^{-1} \text{m}^{-1}$.

\subsection{Self-phase modulation of the pump}
\label{parameters_SPM}
The effect of SPM of the pump is visible in the high gain regime:
the cross-mode TFs show a clear broadening, as well as splitting of
the pump function. Also, the same-mode TFs lose their symmetry, which
is consistent with a chirped pump, as illustrated in figure \ref{spectralPhaseOnUss}.

Assuming that dispersion of the pump is negligible, its evolution along the waveguide is given by the differential equation (\ref{diffEqCompletePump}). Having measured $\beta_{\text{p}}(z=0 , \, \omega)$, the SPM interaction strength $\gamma_\text{SPM}$ is the only fitting parameter left, which determines the evolution of the pump spectral amplitude along the waveguide. To obtain the value of $\gamma_\text{SPM}$, we take the complete signal/idler EOMs (\ref{diffEqCompleteSignal},\ref{diffEqCompleteIdler}), with all the other parameters fitted using the low gain regime measurements described above, and find the value of the SPM interaction strength that best fits the high gain TFs according to the error metric defined in Equation (\ref{error_metric}). The best fit parameter was $\gamma_\text{SPM} = 0.56 \, \text{W}^{-1} \, \text{m}^{-1}$.

\section{Asymmetry in the cross-phase modulation coefficients}
\label{factorof3}
In this appendix we provide an analysis of the way in which waves propagating in a material can experience different nonlinear effects because of their polarization relative to a strong pump.
Suppose we have a pump wave with central frequency $\bar{\omega}_{\rm p}$ and a signal wave with central frequency $\bar{\omega}_{\rm s}$ polarized in the same direction, say the $y$ direction. Then the polarization term responsible for the XPM will  be\cite{boyd2003nonlinear}
\begin{align}
& P_{y}(\bar{\omega}_{\rm s})=\epsilon_{0}\chi_{yyyy}^{(3)}E_{y}(\bar{\omega}_{\rm s})E_{y}(\bar{\omega}_{\rm p})E_{y}(-\bar{\omega}_{\rm p})+\ldots
\end{align}
where $\ldots$ are terms obtained by permuting the different fields appearing in the first term. The fields associated with the three frequencies ($\bar{\omega}_{\rm s}$,
$\bar{\omega}_{\rm p}$, -$\bar{\omega}_{\rm p}$) can be combined in any order, giving
$3!=6$ different terms. So in all we would have 
\begin{align}
& P_{y}(\bar{\omega}_{\rm s})=6\epsilon_{0}\chi_{yyyy}^{(3)}E_{y}(\bar{\omega}_{\rm s})E_{y}(\bar{\omega}_{\rm p})E_{y}(-\bar{\omega}_{\rm p}).\label{eq:same}
\end{align}
Now suppose the pump is in the $y$ direction but the idler is
in the $x$ direction. Then one of the terms contributing to the cross-phase
modulation would be 
\begin{align}
& P_{x}(\bar{\omega}_{\rm i})=\epsilon_{0}\chi_{xxyy}^{(3)}E_{x}(\bar{\omega}_{\rm i})E_{y}(\bar{\omega}_{\rm p})E_{y}(-\bar{\omega}_{\rm p})+\ldots
\end{align}
For this particular $\chi^{(3)}$ component we could put the fields
at $\bar{\omega}_{\rm p}$ and $-\bar{\omega}_{\rm p}$ in other orders, ($2!$ permutations)
so from this particular component we would expect 
\begin{align}
& P_{x}(\bar{\omega}_{\rm i})=2\epsilon_{0}\chi_{xxyy}^{(3)}E_{x}(\bar{\omega}_{\rm i})E_{y}(\bar{\omega}_{\rm p})E_{y}(-\bar{\omega}_{\rm p}).
\end{align}
However, there will be other tensor components involving $x's$ and
$y's$. In particular we can expect a $\chi_{xyyx}^{(3)}$ and a $\chi_{xyxy}^{(3)}$.
In each one of these there will be two permutations of the fields
at $\bar{\omega}_{\rm p}$ and $-\bar{\omega}_{\rm p}$ that share the same Cartesian
component, so in all 
\begin{align}
& P_{x}(\bar{\omega}_{\rm i})=2\epsilon_{0}\chi_{xxyy}^{(3)}E_{x}(\bar{\omega}_{\rm i})E_{y}(\bar{\omega}_{\rm p})E_{y}(-\bar{\omega}_{\rm p})\\
& +2\epsilon_{0}\chi_{xyyx}^{(3)}E_{y}(\bar{\omega}_{\rm p})E_{y}(-\bar{\omega}_{\rm p})E_{x}(\bar{\omega}_{\rm i})\\
& +2\epsilon_{0}\chi_{xyxy}^{(3)}E_{y}(\bar{\omega}_{\rm p})E_{x}(\bar{\omega}_{\rm i})E_{y}(-\bar{\omega}_{\rm p})\\
& =2\epsilon_{0}(\chi_{xxyy}^{(3)}+\chi_{xyyx}^{(3)}+\chi_{xyxy}^{(3)})E_{x}(\bar{\omega}_{\rm i})E_{y}(\bar{\omega}_{\rm p})E_{y}(-\bar{\omega}_{\rm p}).
\end{align}
That is for different polarizations. For the same polarization we
have (\ref{eq:same}). So
\begin{align}
 P_{y}(\bar{\omega}_{\rm s})=&6\epsilon_{0}\chi_{yyyy}^{(3)}E_{y}(\bar{\omega}_{\rm s})E_{y}(\bar{\omega}_{\rm p})E_{y}(-\bar{\omega}_{\rm p})\quad \text{(same pol.)}\\
 P_{x}(\bar{\omega}_{\rm s})=&2\epsilon_{0}(\chi_{xxyy}^{(3)}+\chi_{xyyx}^{(3)}+\chi_{xyxy}^{(3)}) \times \\ &E_{x}(\bar{\omega}_{\rm s})E_{y}(\bar{\omega}_{\rm p})E_{y}(-\bar{\omega}_{\rm p})\quad \text{(diff. pol.)}
\end{align}
If one assumes the following symmetry in the third order susceptibility (satisfied for instance by an isotropic medium) 
\begin{align}
& \chi_{yyyy}^{(3)}=\chi_{xxyy}^{(3)}+\chi_{xyyx}^{(3)}+\chi_{xyxy}^{(3)},
\end{align}
so 
\begin{align}
 P_{y}(\bar{\omega}_{\rm s})&=6\epsilon_{0}(\chi_{xxyy}^{(3)}+\chi_{xyyx}^{(3)}+\chi_{xyxy}^{(3)}) \\
& \times E_{y}(\bar{\omega}_{\rm s})E_{y}(\bar{\omega}_{\rm p})E_{y}(-\bar{\omega}_{\rm p})\quad  \text{(same pol.)}\\
 P_{x}(\bar{\omega}_{\rm i})&=2\epsilon_{0}(\chi_{xxyy}^{(3)}+\chi_{xyyx}^{(3)}+\chi_{xyxy}^{(3)}) \\
 &E_{x}(\bar{\omega}_{\rm i})E_{y}(\bar{\omega}_{\rm p})E_{y}(-\bar{\omega}_{\rm p})\quad \text{(diff. pol.)},
\end{align}
differing precisely a factor of $3$. To the best of our knowledge, the tensor components of the third order nonlinear susceptibility of ppKTP have not been reported in the literature; however, this simple calculation provides a plausible argument for the approximate factor of 3 found between the different cross-phase modulation constants of the signal and idler fields.

\end{document}